\documentclass{svjour3}  
\smartqed  

\usepackage{subfigure,amsmath,amssymb}
\usepackage{graphicx}
\usepackage{color}
\usepackage{graphicx}

\usepackage[colorlinks=true,linkcolor=blue,citecolor=blue]{hyperref}

\newcommand{\be}{\begin{equation}}
\newcommand{\ee}{\end{equation}}
\newcommand{\bea}{\begin{eqnarray}}
\newcommand{\eea}{\end{eqnarray}}

\newcommand{\eref}[1]{Eq.~(\ref{#1})}%
\newcommand{\fref}[1]{Fig.~\ref{#1}} %
\newcommand{\sref}[1]{Sec.~\ref{#1}}%
\newcommand{\aref}[1]{Appendix~\ref{#1}}%
\newcommand{\tref}[1]{Table~\ref{#1}}%

\begin{document}

\title{Shock propagation in a driven hard sphere gas: molecular dynamics simulations and hydrodynamics}      
\titlerunning{Blast waves}
\author{Amit Kumar \and R. Rajesh}

\institute{Amit Kumar \at
              	The Institute of Mathematical Sciences, CIT Campus, Taramani, Chennai 600113, India \\
              	Homi Bhabha National Institute, Training School Complex, Anushakti Nagar, Mumbai 400094, India\\
               	\email{kamit@imsc.res.in}  	    
         \and
             R. Rajesh \at
              	The Institute of Mathematical Sciences, CIT Campus, Taramani, Chennai 600113, India \\
              	Homi Bhabha National Institute, Training School Complex, Anushakti Nagar, Mumbai 400094, India\\
               	\email{rrajesh@imsc.res.in}           
}

\date{Received: \today / Accepted: }

\maketitle

\begin{abstract}
The continuous injection of energy in a stationary gas creates a shock wave that propagates radially outwards. We study the hydrodynamics of this disturbance using event driven molecular dynamics of a hard sphere gas in two and three dimensions, the numerical solution of the Euler equation with a virial equation of state for the gas, and the numerical solution of the Navier-Stokes equation, for the cases when the driving is localised in space and when it is uniform throughout the shock. We show that the results from the Euler equation do not agree with the data from hard sphere simulations when the driving is uniform and has singularities when the driving is localised. Including dissipative terms through the Navier-Stokes equation results in reasonably good description of the data, when the coefficients of dissipation are chose parametrically.

\end{abstract}

\maketitle

\section{\label{sec1-Introduction}Introduction}
The study of shock propagation following an intense explosion is a classic problem in gas dynamics~\cite{landaubook,barenblatt1996scaling,whitham2011linear}. In the initial transient phase, the system emits energy through radiation. However, as the system cools down, it transitions into the hydrodynamic phase, where the primary means of energy transport are the movements of particles, and the significance of radiation diminishes. The disturbance grows radially outward with a shock front separating the affected region from the ambient region. Across this front, the thermodynamic quantities  like density, velocity, temperature, and pressure change abruptly, and the magnitude of these discontinuities is determined by the Rankine-Hugoniot boundary conditions~\cite{landaubook,barenblatt1996scaling,whitham2011linear}. Straightforward dimensional analysis reveals that the radius, $R(t)$, of the shock front exhibits a power-law growth with respect to time $t$ as $R(t)\sim \left(E_i t^2/\rho_0\right)^{{1}/{(d+2)}}$ in $d$-dimensions, where $E_i$ and $\rho_0$ are the initial input energy and ambient mass density of the gas~\cite{landaubook,taylor1950formation,taylor1950formation2,jvneumann1963cw,sedov_book,sedov1946,dokuchaev2002self,stanyukovich2016unsteady}. The  power-law exponent  has been confirmed in the  Trinity explosion~\cite{taylor1950formation,taylor1950formation2}, and in blast waves produced by the deposition of laser pulses in gas jets~\cite{edwards2001investigation}, plasma~\cite{edens2004study}, and atomic clusters of different gases~\cite{moore2005tailored}. 

Beyond the power-law growth of $R(t)$, it is also feasible to obtain the spatio-temporal behavior of density, velocity, temperature, and pressure. These are governed by the continuity equations for mass, momentum, and the energy. In the scaling limit, dissipation factors such as heat conduction and viscosity become negligible, and the hydrodynamics is governed by the Euler equation. For an ideal gas, within the assumption of local equilibrium, Taylor, von-Neumann, and Sedov obtained the exact solution of scaling functions for density, velocity, temperature, and pressure~\cite{taylor1950formation,taylor1950formation2,jvneumann1963cw,sedov_book,sedov1946}.  We refer to this self-similar solution as the TvNS solution. Only recently has the validity of the TvNS theory been checked in simulations of hard spheres in three~\cite{jabeen2010universal,joy2021shock3d,kumar2022blast}, two~\cite{jabeen2010universal,kumar2022blast,joy2021shock,barbier2016microscopic} and one dimensions~\cite{ganapa2021blast,chakraborti2021blast}. It was found that the event driven molecular dynamics (EDMD) simulations in two and three dimensions found significant differences between the TvNS theory and simulations near the shock center and quantified through the  exponents governing the power law behavior of the different scaling functions~\cite{joy2021shock3d,kumar2022blast,joy2021shock}. It was shown that when dissipation terms are included in the Euler equation giving rise to the Navier-Stokes equation (NSE), then the discrepancies of the theory with simulations can be accounted for both in one dimension~\cite{ganapa2021blast,chakraborti2021blast} as well as higher dimensions~\cite{kumar2022blast}. The crossover behavior of the scaling functions from the Euler solution to the Navier-Stokes solution near the shock center has been quantified in one~\cite{ganapa2021blast,chakraborti2021blast}  and two dimensions~\cite{singh2023blast}. This resolution shows that the  order of taking the limits -- first taking the scaling limit and then finding the solution, or finding the solution of the Navier-Stokes equation and then taking the scaling limit -- matter for the final answer. In particular, the boundary conditions satisfied near the shock center are different for the two cases. Recently, for blasts in an inhomogeneous medium, we have identified a critical inhomogeneity parameter at which the Euler equation satisfies these boundary conditions, leading to consistent results across all three approaches (Euler, NSE, and EDMD)~\cite{kumar2025shock}.

A closely related problem is that of shocks that are generated when there is a continuous input of energy in the system by an external source. Now, unlike the problem of single impact discussed above, the system is now driven away from equilibrium due to the constant energy current. This problem has relevance for the study of the motion of interstellar gas due to the effect continuous energy injection by the stellar wind~\cite{avedisova1972formation,falle1975numerical}. Let the source be such that energy increases as $E(t) = E_0 t^\delta$, where $E_0$ and $\delta$ are positive constants. From  dimensional analysis, one obtains that the radius of shock front grows as $R(t)\sim \left(E_0/\rho_0\right)^{{1}/{d+2}}t^{{(2+\delta)}/{(d+2)}}$, in $d$-dimensions~\cite{dokuchaev2002self}. The TvNS solution for the single impact can now be generalized to $\delta\neq 0$. 

Self-similar solutions for driven shocks have been studied for two types of driving mechanisms, which we refer to as central driving and uniform driving. In central driving, energy is continuously input in a fixed localised region around the  shock center, leading to the rate of change of local entropy being zero away from the shock center~\cite{dokuchaev2002self,masuyama2016spherical}. In uniform driving, energy is continuously input uniformly in the region between the shock center and the shock front (which is moving with time)~\cite{dokuchaev2002self,masuyama2016spherical}. In the central driving, the self-similar solution of the Euler equation becomes singular at a finite scaled radius, and thus is unable to describe the hydrodynamics of the shock in the complete region from the shock center to the shock front.  On the other hand, for the uniform driving, there exists a self-similar solution of the Euler equation for the entire region of the shock~\cite{dokuchaev2002self}. The exact self similar solution of the Euler equation for the uniform driving of an ideal gas in three dimensions was found by Dokuchaev~\cite{dokuchaev2002self}. 

In this paper, we focus on the hydrodynamics of the shocks in the presence of an energy source. Given that the correct description of the shock due to a single impact required dissipation terms (Navier-Stokes equation), it is highly likely that the driven shocks also are not described by the Euler equation specially close to the shock center. In addition, it is not even clear whether the Navier-Stokes equation can describe the different thermodynamic quantities for the driven shock, given that the system is far from equilibrium. Also, for central driving, the Euler equation is manifestly insufficient to describe the hydrodynamics, and we ask whether including the dissipation terms to the Euler equation can regularise the singular behaviour seen in the self-similar solution. To address these issues, we study the problem of driven shocks using different approaches namely Euler equation, Navier-Stokes equation, and EDMD simulations for hard sphere gas in two and three dimensions. In two dimensions, we study both central as well as uniform driving, while in three dimensions, we restrict ourselves to only central driving.

The remainder of the paper is organized as follows. In \sref{driventheory}, we review the exact solution of Euler equation for driven shocks in an ideal gas, and find the asymptotic behavior of different scaling functions.  In \sref{nondime_viral}, we modify the equation of state from ideal to virial equation of state to account for steric effects. We then numerically solve the Euler equation and quantify the effect of excluded volume on the solution. In \sref{MDsimulation} we numerically study the driven shock in a hard sphere gas using EDMD simulations, and verify the correctness of the simulations  by benchmarking the known behavior of  physical quantities. In \sref{hydro}, we provide the direct numerical solution (DNS) of the Navier-Stokes equation for virial equation of state, and do a parametric study to understand the effect of the dissipation terms on the scaling functions.  In \sref{comparisonEDMDhydro}, we compare all the results obtained from the theory, EDMD, and DNS of NSE for hard sphere gas. We show that for central driving, the solution for NSE has self-similar solution that spans the entire shock, overcoming the singular behaviour that the solution to the Euler equation suffers from. For both driving, we show that the data from EDMD is well described by NSE. We conclude with a summary and discussion in \sref{summary}.

\section{ \label{driventheory} Review of Euler Equation for Driven Shock}
 In this section, we summarize the Euler equation describing the macroscopic dynamics of a driven shock. Consider a gas at rest having uniform density $\rho_0$, and hence zero pressure and zero temperature everywhere. Energy is isotropically and continuously injected at one point (taken to be the origin) such that the total energy increases with time $t$ as $E(t)=E_0 t^\delta$, $\delta \ge 0$. The driving generates a spherically symmetric shock which expands self similarly in time into the ambient gas. We define the shock front as the surface of discontinuity which separates the moving gas from the ambient stationary gas. A shock is said to be strong if ${p_1}/{\rho_1}\gg {p_0}/{\rho_0}$, where $p$ is the pressure, $\rho$ is the mass density, and subscripts $1$ and $0$ indicate the quantities just behind and front of the shock respectively. Since $p_0=0$ for an initial stationary gas, the shock is always strong.

The scaling of the radius of shock front, $R(t)$, with time $t$, is uniquely determined by  dimensional analysis~\cite{landaubook,sedov_book,stanyukovich2016unsteady}, and in $d$-dimensions is 
\begin{align}
R(t) \sim \left(\frac{E_0}{\rho_0}\right)^{1/(2+d)}t^{(2+\delta)/(2+d)}. \label{frontradius}
\end{align}
The macroscopic state of the gas at time $t$ and position $\vec{r}$ is described in terms of the following fields: density $\rho(\vec{r},t)$, velocity $\vec{v}(\vec{r},t)$, temperature $T(\vec{r},t)$, and pressure $p(\vec{r},t)$. Due to the spherical symmetry, the thermodynamic quantities depend only on radial distance $r$, and the velocity is radial,
\begin{align}
\vec{v}(\vec{r},t)=v(r,t)\hat{r}. \label{radvel}
\end{align}

The continuity equations of locally conserved quantities, mass, momentum, and the energy, give the evolution of the fields. In the scaling limit, $r\to \infty$, $t\to \infty$, such that  $rt^{-(2+\delta)/(2+d)}$ remains constant, the contribution of heat conduction and viscosity become negligible and the hydrodynamics is governed by the  Euler equation. The continuity  equations for mass and momentum along radial direction in $d$-dimensional spherical polar coordinates is given by~\cite{landaubook,barenblatt1996scaling,whitham2011linear,sedov_book,dokuchaev2002self}
\begin{align}
&\partial_t \rho + \partial_r (\rho v) +\frac{(d-1)\rho v}{r} =0,\label{masseq}\\
&\partial_t v + v \partial_r v + \frac{1}{\rho} \partial_r p =0, \label{momentumeq}
\end{align}
The continuity of energy, after assuming the existence of some non-hydrodynamic radiative mechanism of local entropy production, which takes care of the external energy source, is~\cite{dokuchaev2002self}
\begin{align}
&v\left(\epsilon + \frac{p}{\rho} + \frac{v^2}{2}\right) =  \frac{ U r}{R(t)}\left(\epsilon +\frac{v^2}{2}\right),  \label{energyeq}
\end{align}
where $U=\dot{R}(t)$ is the speed of the shock front, and $\epsilon$ is the internal energy per unit volume of the gas. The integral form of the continuity equation for energy in Eq.~(\ref{energyeq}) is obtained from the differential form by integrating over a tiny shell in radial direction. We refer to Ref.~\cite{dokuchaev2002self} for the derivation. For a gas, $\epsilon= T/(\gamma-1)$, where $\gamma$ is the adiabatic constant.  

For central driving, it is convenient to study the continuity equation for entropy:
\begin{align}
&\dot{s}= \partial_t s + v \partial_r s,  \label{entropyeq}
\end{align}
where the local entropy $s=\log\left({p}/{\rho^\gamma}\right)/(\gamma-1)$. For central driving, the flow can be assumed to be adiabatic away from the shock center, i.e.,  $\dot{s}=0$ away from the shock center. 

Equations~(\ref{masseq}-\ref{momentumeq}) and \eref{energyeq} together describe the driven shock problem with uniform driving~\cite{dokuchaev2002self}, while Eqs.~(\ref{masseq}-\ref{momentumeq}) and \eref{entropyeq} with $\dot{s}=0$ together describe the driven shock problem with the central driving~\cite{masuyama2016spherical}.

Assuming local thermal equilibrium, the local pressure $p$ is related to the local temperature $T$ and local density $\rho$ through an equation of state (EOS), reducing the number of variables by one. A general EOS can be written as
\begin{align}
p=k_B\rho T \mathbb{Z}(\rho), \label{generalEOS}
\end{align}
where $\mathbb{Z}(\rho)$ is known as the compressibility factor of the EOS.

Across the shock front these thermodynamic quantities become discontinuous. The values of these quantities ahead and behind the shock front are related by the  Rankine-Hugoniot boundary conditions~\cite{landaubook,whitham2011linear}:
\begin{align}
&\rho_1=\left[1+\frac{2}{(\gamma-1)\mathbb{Z}(\rho_1)}\right]\rho_0,\label{RHdens}\\
&v_1=\frac{2U}{2+(\gamma-1)\mathbb{Z}(\rho_1)},\label{RHvel}\\
&p_1=\frac{2\rho_0 U^2}{2+(\gamma-1)\mathbb{Z}(\rho_1)}.\label{RHpress}
\end{align}
$\gamma=1+{2}/{d}$ is the adiabatic constant for mono-atomic gas in $d$-dimensions.

It should also be noted that total energy $E(t)$ of the gas at time $t$ should be equal to $E_0 t^\delta$, i.e.
\begin{align}
& E_0 t^\delta=\int_0^{R(t)}\left( \frac{\rho v^2}{2}+\frac{\rho T}{\gamma-1}\right) S_d r^{d-1} dr,\label{totalenergy}
\end{align}
where $S_d=2\pi^{d/2} /\Gamma\left(d/2\right)$ is the surface area of $d$-dimensional sphere of unit radius.
The continuity equations~(\ref{masseq})--(\ref{entropyeq}) are first order partial differential equations  in both time and distance. These equations can be converted into ordinary differential equations using self similar solutions.  We define dimensionless distance $\xi$ and non-dimensionalised scaling functions $\widetilde{R}, \widetilde{u},  \widetilde{T},  \widetilde{P}$ corresponding to density $\rho$, velocity $u$, temperature $T$, and pressure $p$ respectively  as
\begin{align}
\xi&=r \left(\frac{E_0}{\rho_0}\right)^{-1/(2+d)}t^{-(2+\delta)/(2+d)}, \label{rescadist} \\
\rho(r,t)&=\rho_0 \widetilde{R}(\xi), \label{rescadens} \\
v(r,t)&=\frac{r}{t}\widetilde{u}(\xi), \label{rescavel}\\
T(r,t)&=\frac{r^2}{t^2}\widetilde{T}(\xi), \label{rescatemp}\\
p(r,t)&=\frac{\rho_0 r^2}{t^2}\widetilde{P}(\xi). \label{rescapres}
\end{align}

We now specialize the solution to the ideal gas for which an exact solution may be found for the case of uniform driving case. For ideal gas $\mathbb{Z}(\rho)=1$, and the equation of state (\eref{generalEOS}) implies that, 
\begin{align}
&\widetilde{P}=\widetilde{R} \widetilde{T}. \label{rescarelideal}
\end{align}
The continuity equations (\ref{masseq}-\ref{energyeq}) and \eref{entropyeq} with $\dot{s}=0$, in terms of the scaling functions, reduce to 
\begin{align}
&(\alpha-\widetilde{u}) \frac{d \log \widetilde{R}}{d \log \xi} - \frac{d \widetilde{u}}{d \log \xi} = d\widetilde{u},\label{idealmass}\\
&(\widetilde{u}-\alpha) \frac{d\widetilde{u}}{d\log \xi} +\widetilde{T}\frac{d \log \widetilde{R}}{d\log \xi} + \frac{d \widetilde{T}}{d\log\xi} +\widetilde{u}[\widetilde{u} -1] +2\widetilde{T} =0,\label{idealmomentum}\\
&\widetilde{T} = \frac{\widetilde{u}^2(\alpha-\widetilde{u})(\gamma-1)}{2(\gamma \widetilde{u}-\alpha))},\label{idealenergy}\\
&(\widetilde{u}-\alpha)\frac{d}{d \log \xi} \log\left( \frac{\widetilde{T}}{{\widetilde{R}}^{\gamma-1}}\right) + 2(\widetilde{u} -1)=0,\label{idealentropy}
\end{align}
respectively, where $\alpha=(2+\delta)/(2+d)$, while the Rankine-Hugoniot boundary conditions reduce to
\begin{align}
&\widetilde{R}(\xi=\xi_f)=\frac{\gamma+1}{\gamma-1},\label{idealRHmass}\\
&\widetilde{u}(\xi=\xi_f)=\frac{2\alpha}{\gamma+1},\label{idealRHpress}\\
&\widetilde{T}(\xi=\xi_f)= \frac{2\alpha^2 (\gamma-1)}{(\gamma+1)^2}, \label{idealRHenergy}
\end{align}
where $\xi_f$ is the location of the shock front. For uniform driving, given the boundary conditions (\ref{idealRHmass})--(\ref{idealRHenergy}) at $\xi_f$, the differential equations (\ref{idealmass})--(\ref{idealenergy}) may be integrated to obtain the scaling functions. However, $\xi_f$ remains indeterminate. The value of $\xi_f$ is uniquely fixed from the non-dimensionalised form of Eq.~(\ref{totalenergy}), 
\begin{align}
S_d \int_0^{\xi_f}\left(\frac{\widetilde{R}\widetilde{u}^2}{2}+\frac{\widetilde{R}\widetilde{T}}{\gamma-1}\right) \xi^{d+1} d\xi=1. \label{idealbeta}
\end{align}

An analytical solution of Eqs~(\ref{idealmass})--(\ref{idealenergy}) with boundary conditions given in Eqs.~(\ref{idealRHmass})--(\ref{idealRHenergy}) is possible~\cite{dokuchaev2002self}. For the completeness of the results, in \aref{appendixA}, we outline the derivation in  three dimensions.

From the exact solution, the behavior of the thermodynamic quantities near the shock center ( $\xi \to 0$) may be derived. These results will be useful for comparison with results from particle based simulations. We find that when $\xi\to 0$,  then $\widetilde{u}\to {\alpha}/{\gamma}$. The asymptotic behavior of $\widetilde{R}$, $\widetilde{u}$, $\widetilde{T}$, and $\widetilde{P}$ near $\xi \to 0$  for uniform driving in $d$-dimensions is
\begin{align}
\widetilde{u} -\frac{\alpha}{\gamma}& \sim \xi^{\frac{2\gamma+d-2}{\gamma-1}} \label{asy_V}\\
\widetilde{R} &\sim \xi^\frac{d}{\gamma-1} \label{asy_G}\\
\widetilde{T} &\sim \xi^{-\frac{2\gamma+d-2}{\gamma-1}} \label{asy_Z},\\
\widetilde{P} &\sim \xi^{-2} \label{asy_P}. 
\end{align}

The exponents of the different non-dimensionalised thermodynamic quantities only depend on $d$, and are independent of $\delta$, while the exponent of $\widetilde{P}$ is a constant. Since the exponents are independent of $\delta$, the power law behavior of thermodynamic quantities remain same as for the case for shocks arising from a single impact~\cite{joy2021shock3d}.

\begin{figure}
\includegraphics[scale=0.4]{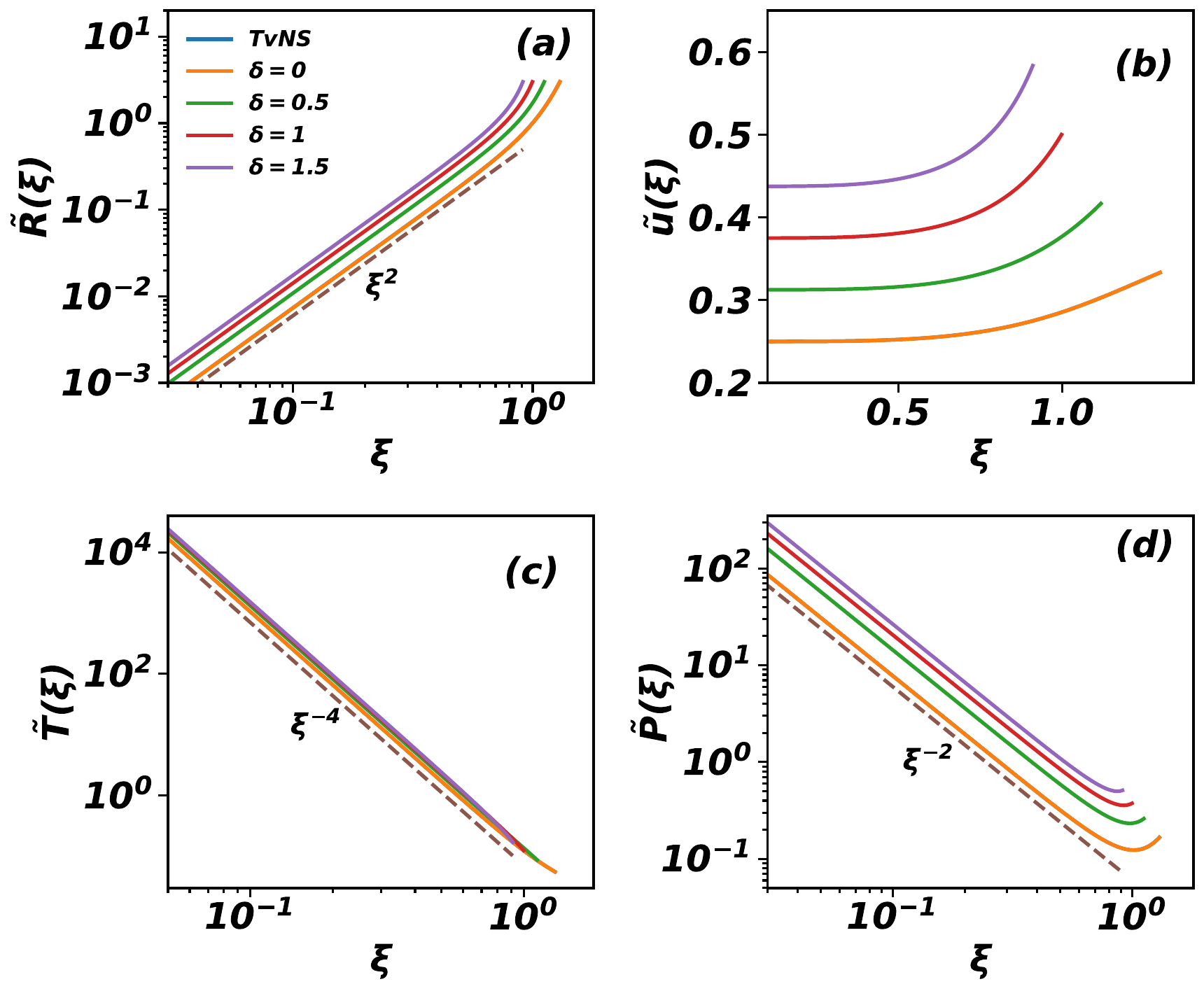}
\caption{The exact solutions of the continuity equations for uniform driving (\ref{idealmass})--(\ref{idealenergy}) for the non-dimensionalised (a) density, $\widetilde{R}$,  (b) velocity, $\widetilde{u}$, (c) temperature, $\widetilde{T}$, and (d) pressure, $\widetilde{P}$  are compared with the asymptotic behavior  in two dimensions when $\xi \to 0$ (see Eqs.~(\ref{asy_V})--(\ref{asy_P})). The data are for four different values of $\delta=0,0.5,1,1.5$. The label $TvNS$ refers to the solution of Euler equation with ideal EOS for a single impact. }
\label{powerlaw_numerical_idealEOS}
\end{figure}

We check for the correctness of the asymptotic analysis for $\xi \to 0$ by comparing them with the full exact solution in two dimensions (see \fref{powerlaw_numerical_idealEOS}), where the non-dimensionalised functions obtained from exact solution of Euler Eqs.~(\ref{idealmass})--(\ref{idealenergy}) are shown for four different values of $\delta=0,0.5,1,1.5$. It is clear that the power laws followed by different thermodynamic quantities are independent of $\delta$, and their exponents are consistent with Eqs.~(\ref{asy_V})--(\ref{asy_P}). Also, we note that the exact solution with $\delta=0$ reproduces the TvNS solution.

\begin{figure}
\includegraphics[scale=0.4]{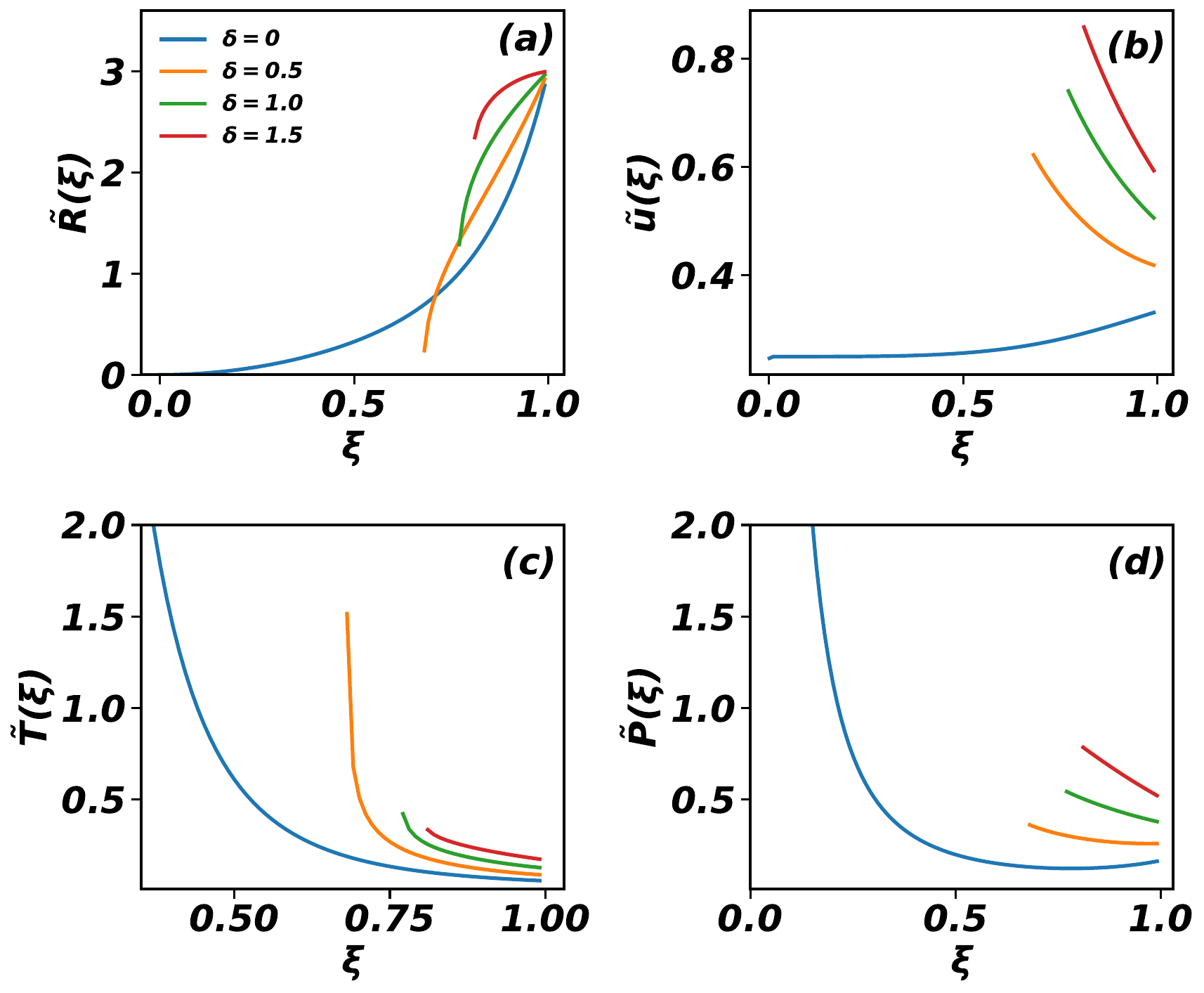}
\caption{The numerical solution of the continuity equations for central driving [Eqs.~(\ref{idealmass}--\ref{idealmomentum}) and Eq.~(\ref{idealentropy})] for the non-dimensionalised (a) density, $\widetilde{R}$,  (b) velocity, $\widetilde{u}$, (c) temperature, $\widetilde{T}$, and (d) pressure, $\widetilde{P}$ for four different values of $\delta=0$, $0.5$, $1$, $1.5$ in two dimensions. The solution curves for different $\delta \neq 0$, do not reach the origin and the range decreases with increasing $\delta$.  } 
\label{Eulercentral}
\end{figure}

We now focus on the case of central driving  (see Eqs.~(\ref{idealmass}--\ref{idealmomentum}) and Eq.~(\ref{idealentropy})) and show that a self-similar solution that spans the entire region of the disturbance  is not possible. Unlike the case of the uniform solution, there is no way to fix $\xi_f$. Hence we fix $\xi_f=1$ for illustrative purposes. In \fref{Eulercentral}, we show the numerical solution of Eqs.~(\ref{idealmass}--\ref{idealmomentum}) and Eq.~(\ref{idealentropy}) in two dimensions. For $\delta \neq 0$, the solution is valid  for only  limited range of $\xi$, and the width of this range of $\xi$ decreases with  increasing $\delta$.

To compare the results with EDMD simulations, we need to take into account steric effects. We describe below the details of how we incorporate excluded volume effects.

\subsection{\label{nondime_viral}Euler equation for hard sphere gas} 

In a hard sphere gas, steric effects are important unlike in ideal gas. Thus, a  more realistic EOS is needed to account for these effects. Virial EOS is the most common EOS for hard spheres, which take the following form,
\begin{align}
p=\rho k_B T \left( 1+ \sum_{i=2}^\infty B_i \rho^{i-1} \right), \label{virialEOS}
\end{align}
with compressibility factor,
\begin{align}
\mathbb{Z}(\rho)= 1+ \sum_{i=2}^\infty B_i \rho^{i-1}. \label{comprevirialEOS}
\end{align}
where $B_i$ denotes the $i^{th}$ virial coefficient. We tabulate the known  values of the virial coefficients~\cite{mccoy2010advanced} in \tref{table1}. For hard core gases $B_i$, does not depend on temperature and is therefore only depends on the shape of the particles. 

\begin{table}
\caption{\label{table1} 
The numerical values of the virial coefficients $B_i$ for two and three dimensional hard sphere gas of particles of diameter one. The data are taken from Ref.~\cite{mccoy2010advanced}.}
\begin{tabular}{lll}
 \hline\noalign{\smallskip}
$i$  & $B_i(d=2)$ & $B_i(d=3)$  \\[0.5ex]
 \noalign{\smallskip}\hline\noalign{\smallskip}
 $2$ & $\frac{\pi}{2}$ & $\frac{2\pi}{3}$\\
 $3$ & $ (\frac{4}{3}-\frac{\sqrt{3}}{\pi}) B_2^2$ & $\frac{5}{8}B_2^2$\\
 $4$ & $\Big[2-\frac{9 \sqrt{3}}{2 \pi}+\frac{10}{\pi ^2}\Big] B_2^3$ & $\Big[\frac{2707}{4480} + \frac{219\sqrt{2}}{2240\pi} - \frac{4131}{4480}\frac{\arccos[1/3]}{\pi} \Big]B_2^3$\\ 
 $5$ & $0.33355604 B_2^4$ & $0.110252B_2^4$\\
 $6$ & $0.1988425 B_2^5$ & $0.03888198B_2^5$\\
 $7$ & $0.11486728 B_2^6$ & $0.01302354B_2^6$\\
 $8$ & $0.0649930 B_2^7$ & $0.0041832B_2^7$\\
 $9$ & $0.0362193 B_2^8$ & $0.0013094B_2^8$\\
 $10$ & $0.0199537 B_2^9$ & $0.0004035B_2^9$\\[1ex]
 \noalign{\smallskip}\hline
\end{tabular}
\end{table}

The Euler Eqs.~(\ref{masseq})--(\ref{energyeq}) with the  hard sphere gas EOS can be simplified in terms of scaling functions as 
\begin{align}
&(\alpha-\widetilde{u}) \frac{d \log \widetilde{R}}{d \log \xi} - \frac{d \widetilde{u}}{d \log \xi} = d\widetilde{u},\label{virialmass}\\
&(\widetilde{u}-\alpha)\xi \frac{dV}{d \xi} +\frac{d(\widetilde{T}\mathbb{Z})}{d\log \xi} + \widetilde{T}\mathbb{Z} \left(\frac{d \log \widetilde{R}}{d \log \xi}\! + \! 2 \!\right)+ \widetilde{u}^2-\widetilde{u} =0,\label{virialmom}\\
&\widetilde{T} = \frac{\widetilde{u}^2(\alpha-\widetilde{u})(\gamma-1)}{2\left[(\gamma-1) \widetilde{u}\mathbb{Z}-(\alpha-\widetilde{u})\right]},\label{virialenergy}
\end{align} 
The Rankine-Hugoniot boundary conditions, Eqs.~(\ref{RHdens})--(\ref{RHpress}),  for the hard sphere gas in terms of scaling functions reduce to
\begin{align}
&\widetilde{R}(\xi_f)=1+\frac{2}{(\gamma-1)\mathbb{Z}},\label{virialRHmass}\\
&\widetilde{u}(\xi_f)=\frac{2\alpha}{2+(\gamma-1)\mathbb{Z}},\label{virialRHvel}\\
&\widetilde{T}(\xi_f)= \frac{2\alpha^2(\gamma-1)}{\left[2+(\gamma-1)\mathbb{Z}\right]^2}.\label{virialRHpress}
\end{align}
The ordinary differential equations (\ref{virialmass})--(\ref{virialenergy}) with the boundary conditions, Eqs.~(\ref{virialRHmass})--(\ref{virialRHpress}) can be solved numerically. As  for the ideal gas, we find the value of $\xi_f$ recursively by satisfying the energy constraint \eref{idealbeta}. 

We now present the solution of Euler equation (see Eqs.~(\ref{virialmass})--(\ref{virialenergy})) only for the case of uniform driving in two dimensions. We obtain the numerical solution of Euler equation for hard disk gas with ambient density $\rho_0 =0.382$, for different values of $\delta$, with virial EOS truncated at different terms. From the numerical solution, we calculated the values of $\xi_f$ when virial EOS truncated at various terms. These values are tabulated in \tref{table2}. We find that $\xi_f$ does not change much between the equation of state truncated at the 8-th and 10-th virial terms for all values of $\delta$.
\begin{table} 
\caption{\label{table2} 
The numerical values of $\xi_f$ for the hard sphere gas when virial EOS truncated at the $i$-th term, in three dimensions. The data for $\xi_f$ for TvNS solution in two dimensions is taken from Ref.~\cite{joy2021shock}. These data are for hard disk gas with diameter one, $\gamma=2$, density $0.382$.} 
\begin{tabular}{llll}
 \hline\noalign{\smallskip}
$i$  & $\xi_f(TvNS)$ & $\xi_f(\delta=0.5)$ &$\xi_f(\delta=1.0)$  \\[0.5ex]
 \noalign{\smallskip}\hline\noalign{\smallskip}
 $2$ & $1.5564$ & $1.3426$ & $1.1993$\\
 $4$ & $1.7286$ & $1.4904$ & $1.3308$\\ 
 $6$ & $1.7643$ & $1.5202$ & $1.3569$\\
 $8$ & $1.7719$ & $1.5264$ & $1.3623$\\
 $10$& $1.7736$ & $1.5277$ & $1.3634$\\[1.0ex]
 \hline\noalign{\smallskip}
\end{tabular}
\end{table}

\begin{figure}
\includegraphics[scale=0.4]{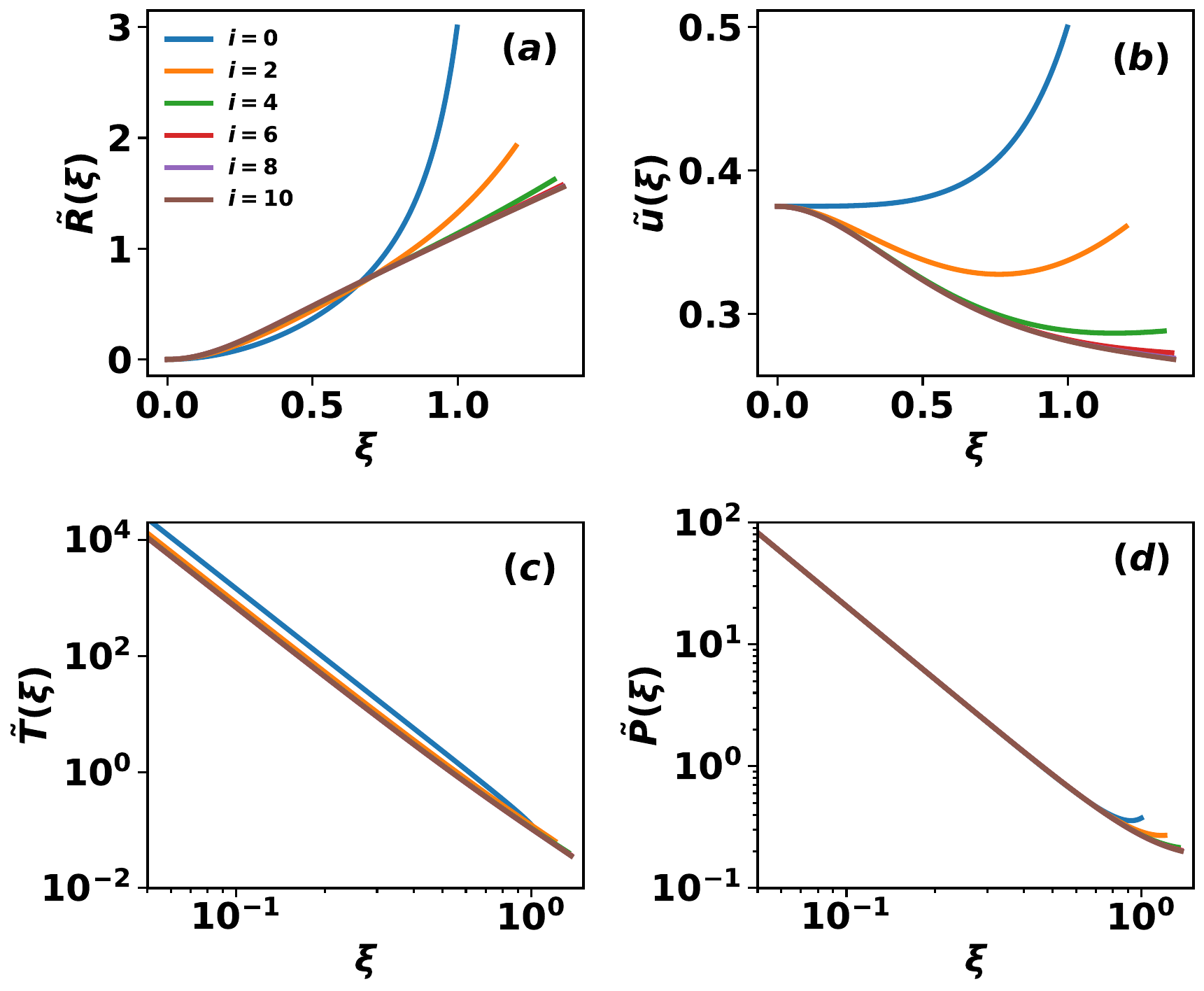}
\caption{Non-dimensionalised thermodynamic quantities obtained by numerically solving Euler equation for uniform driving with virial equation of state (see Eqs.~(\ref{virialmass})--(\ref{virialenergy})) for $\delta=1$ when the virial EOS is truncated at $i=0$,$2$,$4$,$6$,$8$, and $10$-th term. The curves corresponding to $i=8,10$ collapse on each other showing negligible truncation error at $i=10$. $i=0$ represents the ideal EOS. The data are for ambient gas density $\rho_0=0.382$ and $d=2$.}
\label{numerical_k1_virial_term}
\end{figure}

We now examine the role of the truncation of the equation of state on the thermodynamics quantities. Figure~\ref{numerical_k1_virial_term} shows the variation of  the scaling functions $\widetilde{R}$, $\widetilde{u}$, $\widetilde{T}$, and $\widetilde{P}$ with $\xi$ for hard spheres for $\delta=1$ in two dimension, when the virial EOS is truncated at $i=0$, $2$, $4$, $6$, $8$, $10$-th terms.  We find that including the virial terms does affect the thermodynamic quantities, especially density and velocity. However,  the data corresponding to $i=8, 10$ lie on top of each other showing negligible truncation error at $i=10$. Thus, truncating virial EOS at $i=10$ is a good approximation to the actual EOS. We also point out that the exponents characterizing the power-law behavior do not depend on the truncation. 

\begin{figure}
\includegraphics[scale=0.4]{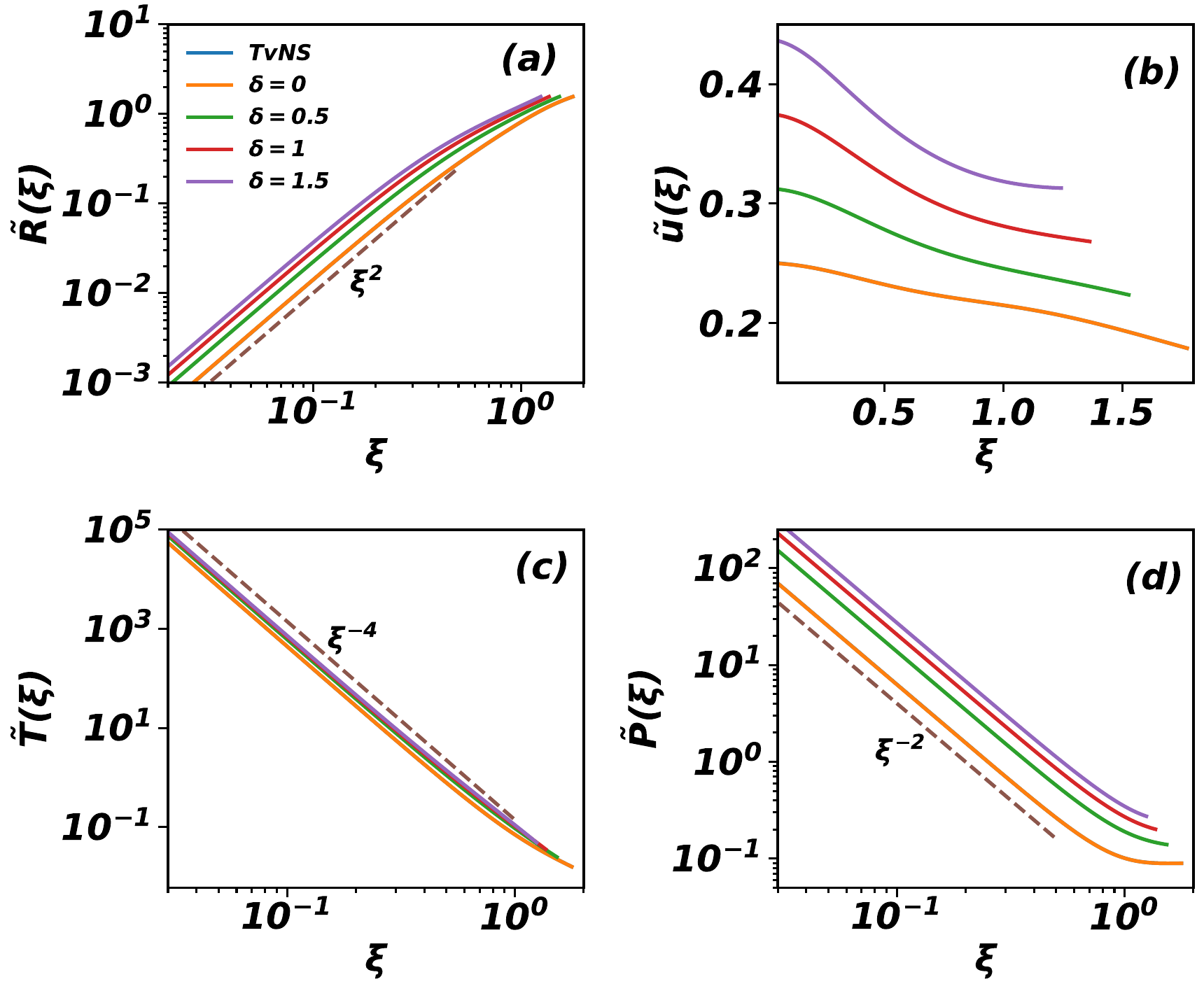}
\caption{Power law behavior of scaling functions for uniform driving obtained from numerical solution of Euler equation for hard spheres (see Eqs.~(\ref{virialmass})--(\ref{virialenergy})) for four different values of $\delta=0,0.5,1,1.5$ in two dimensions. From the plot, power law behaviors of different scaling functions seem to be independent of the value of $\delta$ and are the same as in the ideal gas case. The data shown here are for ambient gas density $\rho_0=0.382$, and for the virial EOS truncated at $i=10$.}
\label{powerlaw_numerical_virialEOS}
\end{figure}

We now examine the role of $\delta$, the driving rate, on the thermodynamic quantities, within the Euler equation. For this, we keep the truncation of the virial expansion fixed at $i=10$ and vary $\delta$. We find that the exponents characterizing the power law behavior of the different thermodynamic quantities are independent of $\delta$ (see  \fref{powerlaw_numerical_virialEOS}), and hence same as that for the single impact with ideal gas EOS. 

\section{\label{MDsimulation} EDMD Simulations}

In this section we briefly describe the details of the EDMD simulations of driven shock in a particle based model. The simulation results are for $\delta=1$ when energy is input at a constant rate, i.e., $E(t)=E_0 t$. 

We first describe the model. Consider a system of $N$ identical hard spheres, labeled $1, 2, \ldots, N$, distributed uniformly in space. The particles are initially at rest. Depending upon the driving scheme, we input of energy at a constant rate either at the  origin (central driving) or throughout the disturbed region (uniform driving).  The system evolves in time through momentum and energy conserving binary collisions between particles. All masses and distances are measured in terms of the particle mass $m$ and diameter $D$, hence we set the mass and diameter of each particle to $1$. Time is measured in terms of the inherent time scale $\left(m D^2/E_0\right)^{1/3}$.   

In a binary collision, the normal component of the relative velocity is reversed while the tangential component remains unchanged. If $\vec{v}_i,\vec{v}_j$ are the pre-collision velocities of colliding particles $i,j$, then their post-collision velocities $\vec{v}'_i,\vec{v}'_j$ are given by
\begin{align}
&\vec{v}'_i = \vec{v}_i - (\hat{n}.\vec{v}_{ij})\hat{n},    \label{colveli}\\
&\vec{v}'_j = \vec{v}_j - (\hat{n}.\vec{v}_{ji})\hat{n},      \label{colvelj}
\end{align}
where $\hat{n}$ is the unit vector along the line joining the centers of the two particles at the time of contact, and $\vec{v}_{ij} = \vec{v}_{i}-\vec{v}_{j}$ is their relative velocity. 

We model the continuous driving as follows. For central driving, consider a sphere of radius $R_0$ centered about the origin or equivalently center of the simulation box. In each time interval $\Delta t$, a particle within the sphere of radius $R_0$ is chosen at random and its velocity is modified to
\begin{align}
&\vec{v}'_i=\vec{v}_i+\vec{\eta},\label{drivevel}
\end{align}
where the components of the noise $\vec{\eta}$ are drawn from a uniform distribution between $-\sqrt{d E_0 \Delta t/6}$ to $\sqrt{d E_0 \Delta t/6}$.
For  uniform driving, after each time interval $\Delta t$, a moving particle is randomly selected among all the moving particles at that time and it's velocity is modified as \eref{drivevel}.

With such a driving it is straightforward to show that  total energy increases as
\begin{align}
E(t)=E_0 t.
\end{align}

The simulations are done using the event driven molecular dynamics scheme where the system evolves from event to event, the events being collisions, driving and cell crossing~\cite{rapaport2004art}. Boundary effects are avoided by choosing the number of particles and box size such that the shock does not reach the boundary within the simulation time. The EDMD simulations were performed  for $N=4 \times 10^7$ particles with mass density $\rho_0=0.4013$, $E_0=2.5 \times 10^{-6}$, and $R_0=15.0$  in three dimensions, and with $N=8 \times 10^6$ particles with mass density $\rho_0=0.382$, $E_0=3.3 \times 10^{-4}$, and $R_0=30.0$ in two dimensions.

We measure the radial density, velocity, temperature, and pressure in our simulation. We define density $\rho(\vec{r},t)$ and velocity $v(\vec{r},t)$ as local average of density and radial velocity at position $\vec{r}$ and time $t$. Local temperature $T(\vec{r},t)$ is defined as the variance of local velocity. We measure local pressure for $d$-dimensional hard spheres~\cite{isobe2016hard} as
 \begin{align}
 p = \rho T - \frac{\rho}{d N'\Delta t'} \sum_{collisions} \vec{r}_{ij}.\vec{v}_{ij},  \label{edmdpress}
 \end{align}
where $\vec{r}_{ij}=\vec{r}_i-\vec{r}_j$ is the distance between the colliding sphere, $\Delta t'$ and $N'$ are time interval and average number of particles belonging to a particular radial bin in which pressure is being measured. 

\begin{figure}
    \centering
    \subfigure[]{\includegraphics[scale=0.18]{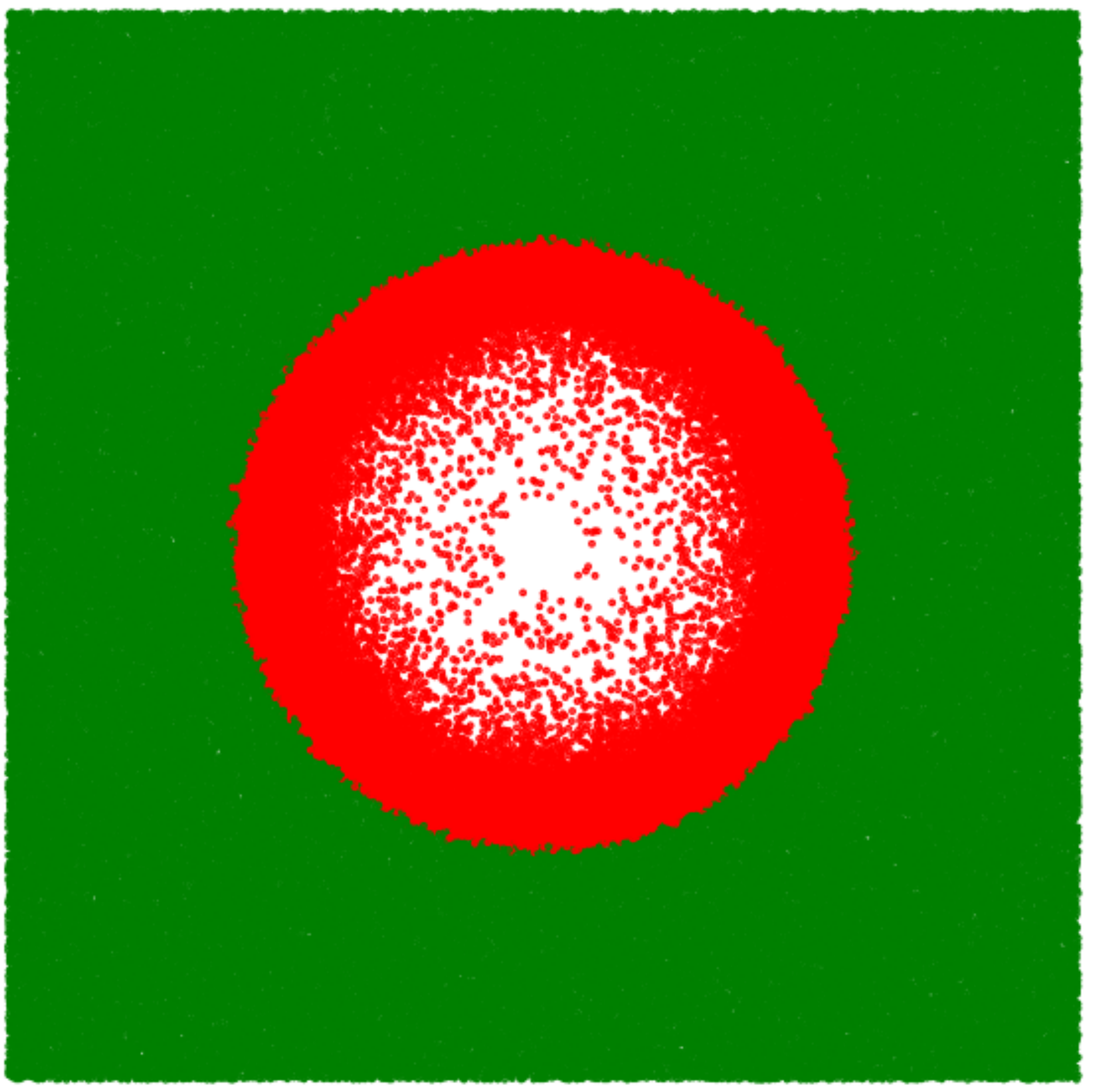}}
    \subfigure[]{\includegraphics[scale=0.18]{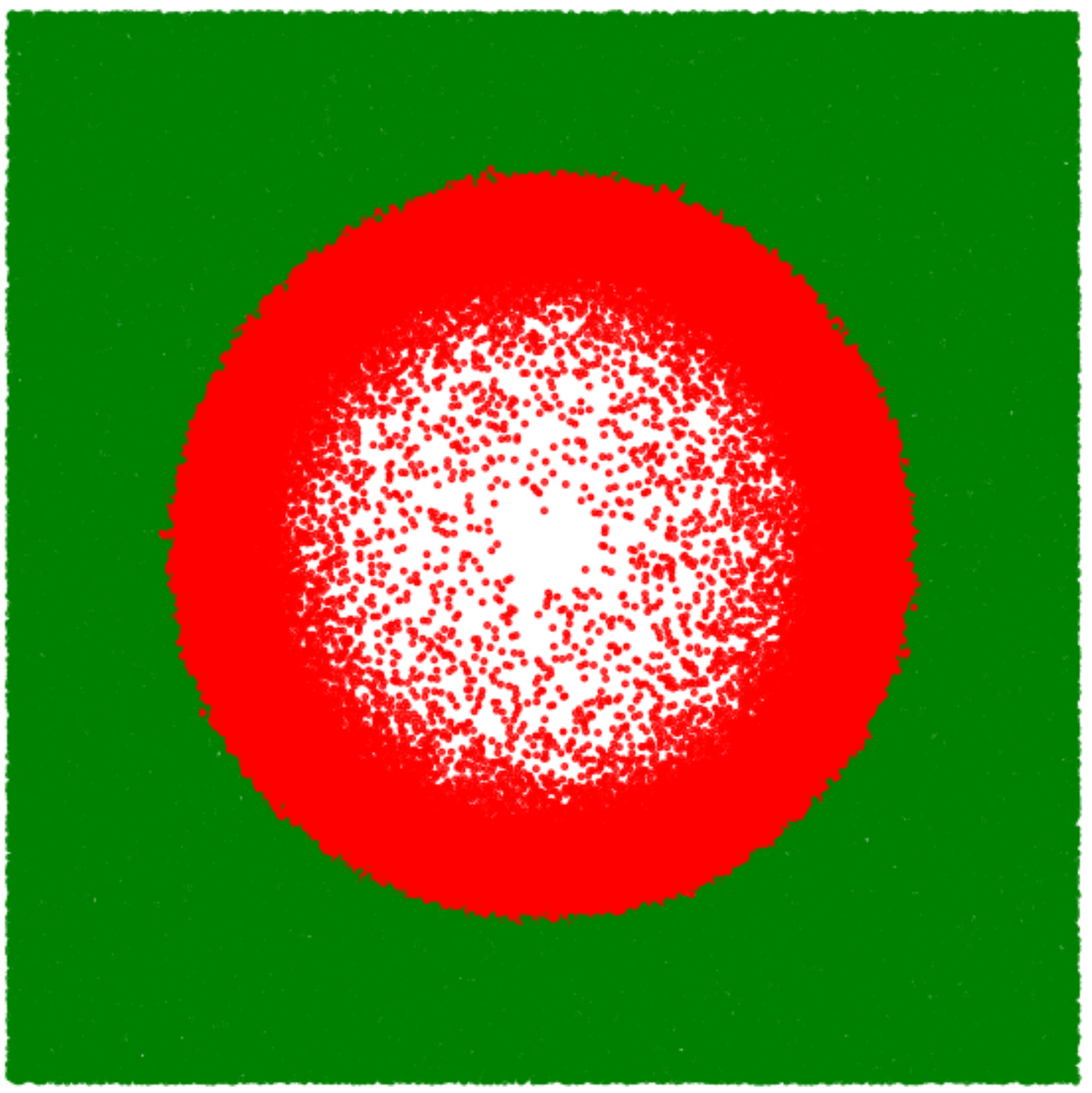}}
    \subfigure[]{\includegraphics[scale=0.18]{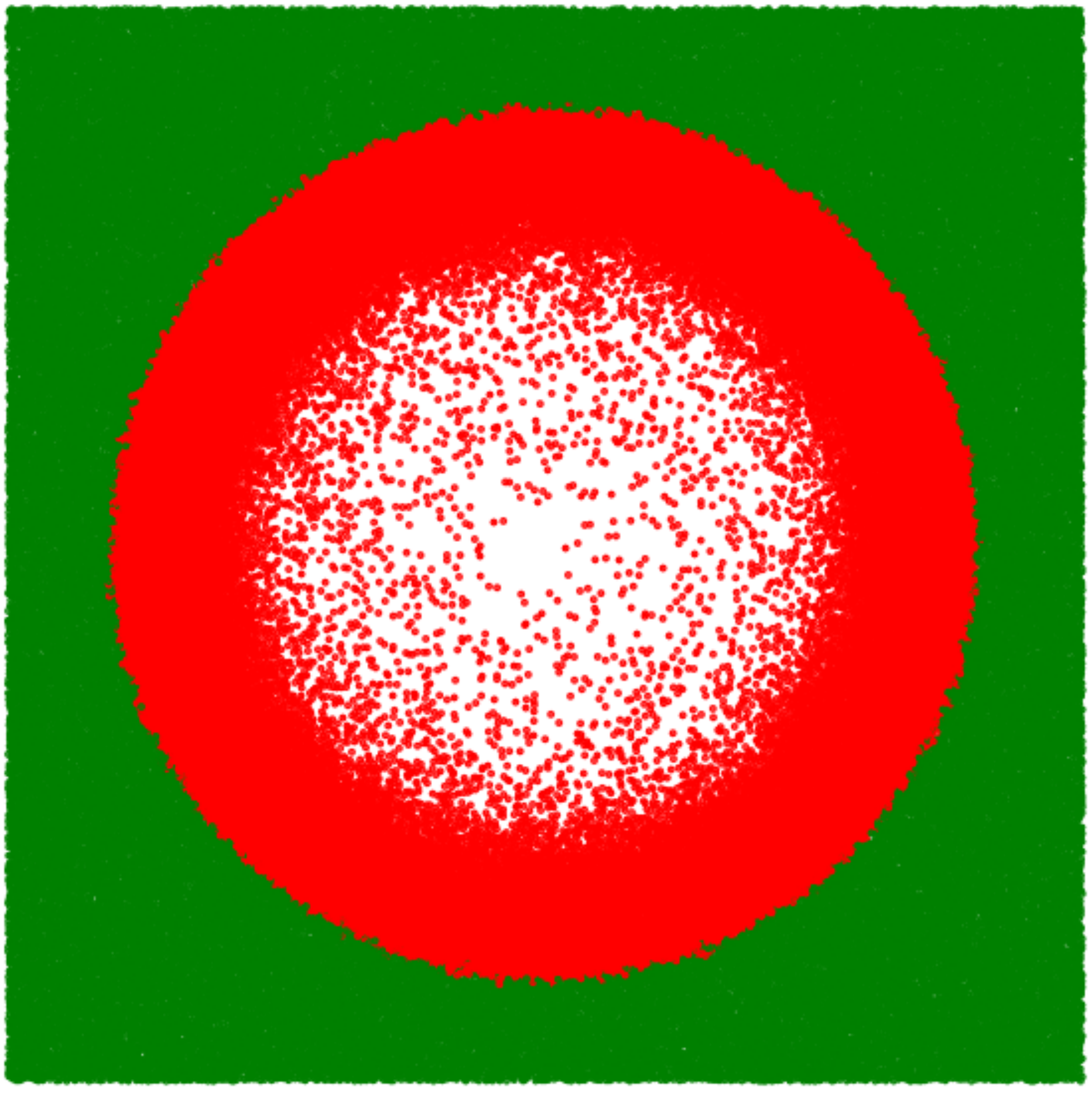}}
    \subfigure[]{\includegraphics[scale=0.18]{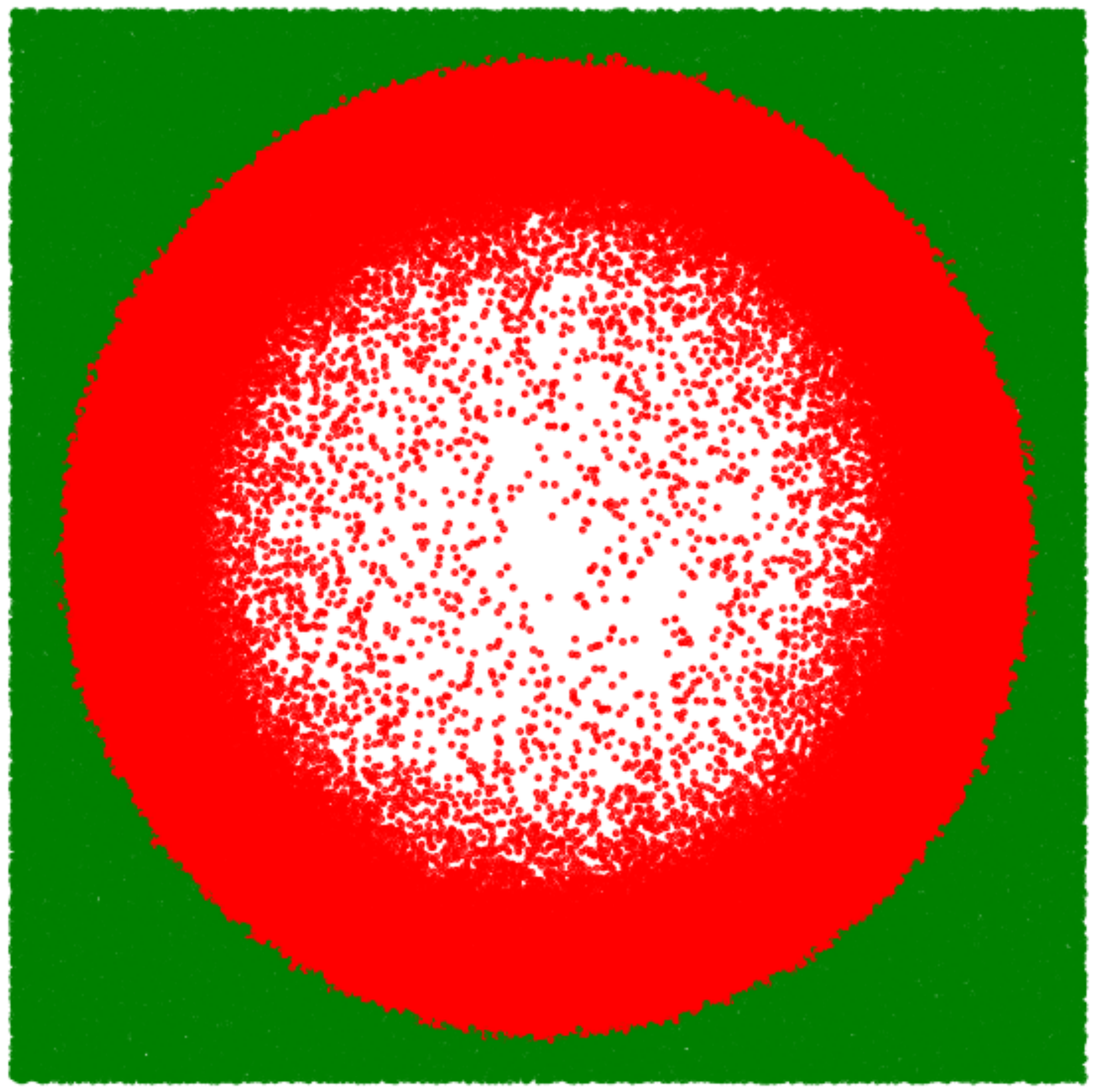}}
    \caption{ Snapshots of a crossection of shock in the $x$-$y$ plane obtained by plotting the coordinates of only the particles with $z$-coordinates between $-1$ to $1$. The data are for the times (a) $t=1357$, (b) $t=1900$, (c) $t=2443$, (d) $t=2986$. Stationary particles are colored green while moving particles are colored red. The data shown here are for ambient gas density $\rho_0=0.4013$ and $2\times 10^7$ number of particles, and for central driving.}
\label{structre}
\end{figure}

The isotropic driving generates a spherically symmetric shock which grows radially outwards. To visualize the growing shock, in \fref{structre}, we plot the $x$- and $y$- coordinates of all the particles lying between the planes $z=-1$ and  $z=1$  for the central driving case in three dimensions. It can be observed that there is a sharp boundary between the moving particles (red) and the stationary particles (green), and the shock front expands in time.  Also, the density near the shock center is close to zero. Similar features are seen for the case of uniform driving also.

\begin{figure}
\includegraphics[scale=0.24]{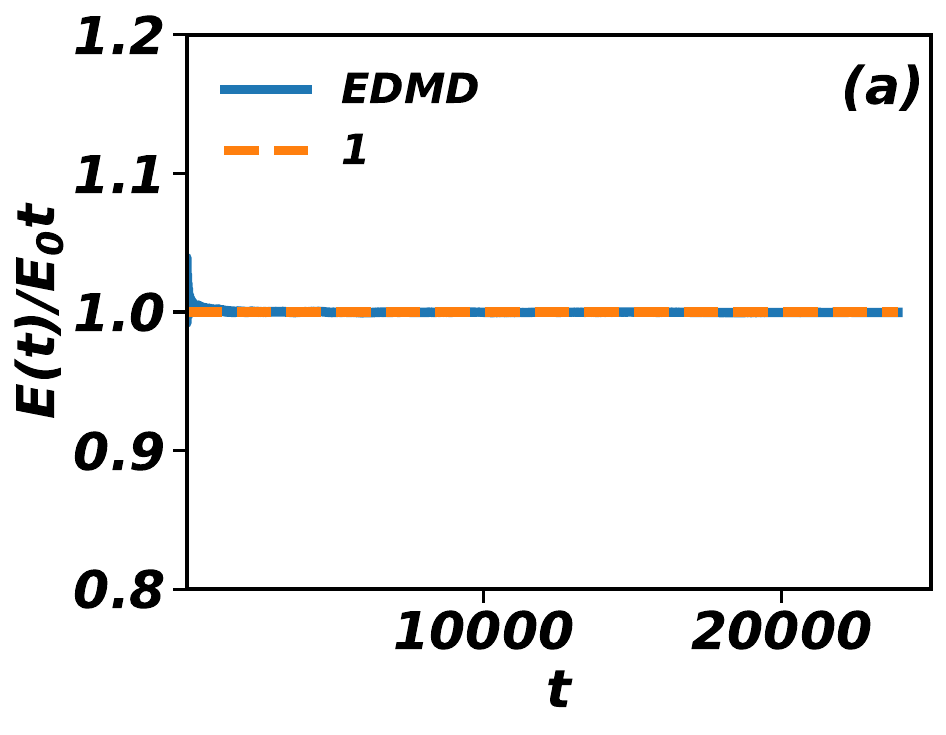}
\includegraphics[scale=0.24]{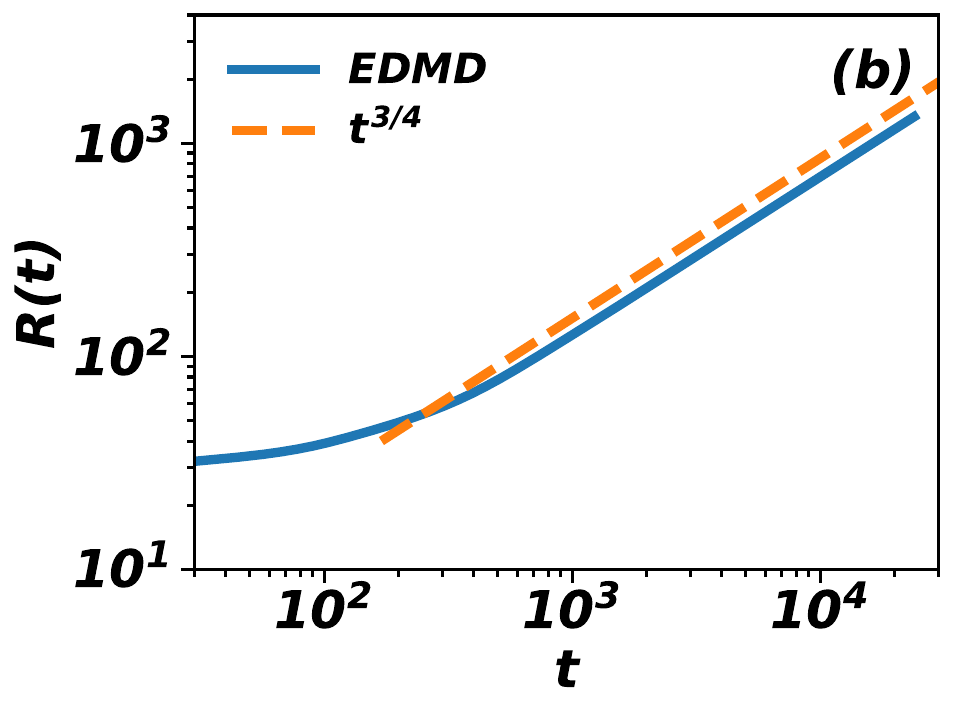}
\includegraphics[scale=0.24]{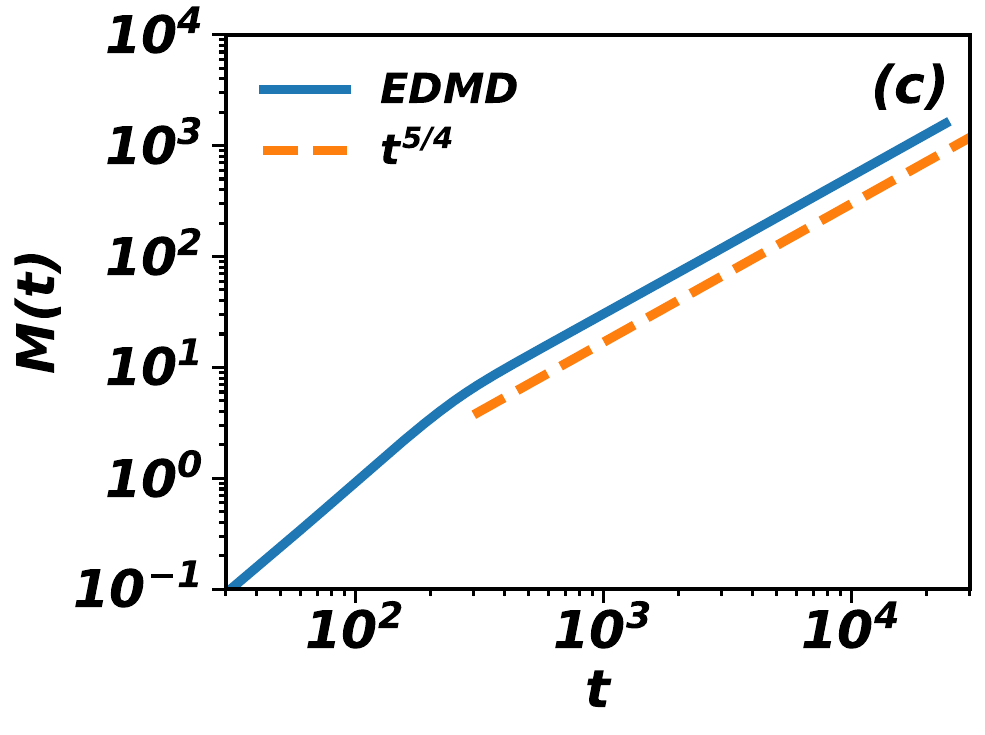}
\caption{Power law growth of (a) total energy $E(t)=E_0 t$, (b) radius of the shock front $R(t)\sim t^{3/4}$, and (c) total radial momentum $M(t)\sim t^{5/4}$ of the system. Solid lines represent the results from EDMD, and dashed lines represent the respective power laws. The results are for the central driving in two dimensions with $\delta=1$.}
\label{powerscaling}
\end{figure}

To benchmark our EDMD simulation, we first confirm that the total energy increases as $E_0 t$, as can be seen from \fref{powerscaling}(a). To further benchmark our simulations, we compare the power law growth of the radius of the shock and the radial momentum with time with known scaling laws. For the  driven shock, in the scaling regime, the total radial momentum $M(t)$, and the radius of shock front $R(t)$ should increase with time as
\begin{align}
M(t) &\sim t^{\alpha(d+1)-1}, \label{eq:totalmomentum}\\
R(t) &\sim t^{\alpha}. \label{eq:totalradius}
\end{align}
The simulation results reproduce these power laws for large time, as can be seen in \fref{powerscaling}(b) and (c). $R(t)$ is measured as the mean value of the radial distance of the moving particles, while  $M(t)$  is obtained as the cumulative radial velocity. For short times, there is a deviation from these power laws for the case of central driving. This is due to the radius of the shock being comparable to the driving scale $R_0$. The crossover time also gives us a measure of the time beyond which the scaling regime is reached. However, for uniform driving, the driving radius is $R_0=R(t)$, so there is no such crossover and the power law behaviors are achieved from the start of the simulation itself.

Before a detailed analysis of the behaviour of the different thermodynamic quantities obtained from EDMD, we first describe the driven shock using Navier-Stokes equation i.e. inclusion of heat conduction and viscosity effects in Euler equation.

\section{\label{hydro} Navier-Stokes equation} 

In Euler equation, it was assumed that at long time, the contribution of the dissipation terms (heat conduction and viscosity) become negligible in the scaling limit.  Since we anticipate the need for dissipation terms to describe the EDMD results, we now include the dissipation terms and describe how we numerically solve the resulting Navier-Stokes equation.

The continuity equations of mass, momentum, and energy, after including the dissipation terms, in the radial coordinates are given by~\cite{landaubook,whitham2011linear,huang1963statistical,warsi2005fluid}
\begin{align}
&\partial_t\rho + \frac{1}{r^{d-1}}\partial_r(r^{d-1}\rho u) = 0,\label{nse_mass}\\
&\partial_t(\rho u) +\frac{1}{r^{d-1}}\partial_r \left[r^{d-1}\rho u^2\right] +\partial_r p  = \frac{1}{r^{d-1}} \partial_r(2\mu r^{d-1} \partial_r u) - \frac{2\mu (d-1)u}{r^2} \nonumber\\
&\quad + \partial_r\left[ \left( \zeta -\frac{2}{d} \mu \right) \frac{1}{r^{d-1}} \partial_r(r^{d-1} u) \right],\label{nse_momentum}\\
&\partial_t\Big(\frac{1}{2}\rho u^2 +  \frac{\rho T}{\gamma-1}\Big) + \frac{1}{r^{d-1}}\partial_r\Big(r^{d-1}\Big[\frac{1}{2}\rho u^2 + \frac{\rho T}{\gamma-1} + p \Big]u\Big) = \frac{1}{r^{d-1}}\partial_r\Big(2 r^{d-1}\mu u  \partial_r u \Big)\nonumber\\
&\quad + \frac{1}{r^{d-1}}\partial_r\left(r^{d-1} u \left[\zeta-\frac{2}{d} \mu \right]  \frac{1}{r^{d-1}}\partial_r\left(r^{d-1} u\right)\right) + \frac{1}{r^{d-1}}\partial_r\left(r^{d-1}\lambda \partial_r T \right) \nonumber\\
&\qquad + {\rm driving~term}, \label{nse_energy}
\end{align}
where $\mu$ is the viscosity, $\lambda$ is the heat conductivity, and $\zeta$ is the bulk viscosity.

The viscosity $\mu$ and heat conduction $\lambda$ of a fluid of hard spheres increase with temperature $T(r,t)$ as~\cite{huang1963statistical,reif2009fundamentals},
\begin{align}
\mu &= C_1 \sqrt{T},\label{eq4}\\
\lambda &=C_2 \sqrt{T}, \label{eq5}
\end{align}
where $C_1$ and $C_2$ are the coefficients of viscosity and heat conduction respectively. From kinetic theory of gases, the approximate values of $C_1$ and $C_2$ for hard sphere particles of diameter $D$, which we denote by  $C_1^*$ and $C_2^*$, are given by~\cite{reif2009fundamentals}
\begin{align}
C^*_1 &= \frac{1}{dD^{d-1}}\sqrt{\frac{mk_B}{\pi^{d-1}}}~ \frac{\Big[ \Gamma\Big(\frac{d+1}{2}\Big)\Big]^2}{ \Gamma\Big(\frac{d}{2}\Big)},\label{eq6}\\
C^*_2 &= \frac{1}{2D^{d-1}}\sqrt{\frac{k_B^3}{m\pi^{d-1}}}~ \frac{\Big[ \Gamma\Big(\frac{d+1}{2}\Big)\Big]^2}{ \Gamma\Big(\frac{d}{2}\Big)},\label{eq7}
\end{align}
where $m$ is the mass of a particle, and $k_B$ is Boltzmann constant.  $\Gamma$ is the Gamma function. The bulk viscosity for mono-atomic gas is zero~\cite{boukharfane2019role}.

We use MacCormack method~\cite{maccormack1982numerical} to numerically solve the Navier-Stokes Eqs.~(\ref{nse_mass})--(\ref{nse_energy}), for $\delta=1$. This method has accuracy up to second order both in time discretization $\Delta t$ and radial discretization $\Delta r$.  We call the numerical solution of NSE as direct numerical solution. The initial conditions on thermodynamic quantities at $t=0$ are given by: constant density everywhere, zero velocity everywhere, and zero temperature everywhere. 
 
For the energy source at the origin for central driving, instead of taking a delta function energy source, we take it as a Gaussian to avoid numerical difficulties. We replace the driving term in Eq.~(\ref{nse_energy}) by 
\begin{equation}
{\rm driving~term}= \frac{A_0 \delta t^{\delta-1}}{S_d r^{d-1}} \exp\left[ \frac{-r^2 A_0^2 \pi}{4 E_0^2} \right]. \label{gaussianinjection}
\end{equation}
For uniform driving, we model the driving term as
\begin{equation}
{\rm driving~term}= \frac{A_0 \delta t^{\delta-1}}{V_d R(t)^d} S_\delta(r, R(t)), \label{uniforminjection}
\end{equation}
where $R(t)$ is the radius of shock front, and $V_d=\pi^{d/2}/\Gamma\left(1+d/2\right)$ is the volume of $d$-dimensional sphere of unit radius. $S_\delta(x,y)$ is the step function defined as,
\begin{align}
& S_\delta(x,y) = \left\{
\begin{array}{ll}
      1 & x \le y, \\
      0 & x > y. \\
\end{array} \right.
\end{align}
Such energy sources lead the total energy of the system to increase as $E_0 t^\delta$. 

To avoid edge effects, we 
choose the system size $L$ in such a manner that shock does not reach to the boundary upto the maximum time we integrate.  We use boundary conditions where at the shock center, $r=0$, the radial derivative of density and temperature are zero, and radial velocity is set to zero, and at the boundary of the region, the initial ambient values are maintained for each of the thermodynamic quantities~\cite{kumar2022blast}. The numerical values of the parameters that we use in our DNS are tabulated in the \tref{table3}. 

\begin{table} 
\caption{\label{table3} 
The numerical values of different parameters used in solving the Navier-Stokes Eqs.~(\ref{nse_mass})--(\ref{nse_energy}).} 
\begin{tabular}{lll}
\hline\noalign{\smallskip}
Parameters  & Values ($d=2$) & Values ($d=3$)  \\[0.5ex]
 \noalign{\smallskip}\hline\noalign{\smallskip}
 $\delta$ & $1.0$ & $1.0$ \\  
 $\Delta r$ & $0.05$ & $0.08$ \\
 $\Delta t$ & $10^{-4}$ & $10^{-4}$  \\
 $A_0$ & $10^{-4}$ & $10^{-4}$  \\
 $\gamma$ & $2$ & $5/3$  \\
 $\rho_0$ & $0.382$ & $0.4013$ \\ 
 $L$ & $1000$ & $300$  \\ 
 $\zeta$ & $0$ & $0$ \\
 $C^*_1$ & $\sqrt{\pi}/8$ & $2/(3\sqrt{\pi^3})$ \\
 $C^*_2$ & $\sqrt{\pi}/8$ & $1/\sqrt{\pi^3}$\\[1ex]
 \hline\noalign{\smallskip}
\end{tabular}
\end{table}

\begin{figure}
\centering
\includegraphics[scale=0.24]{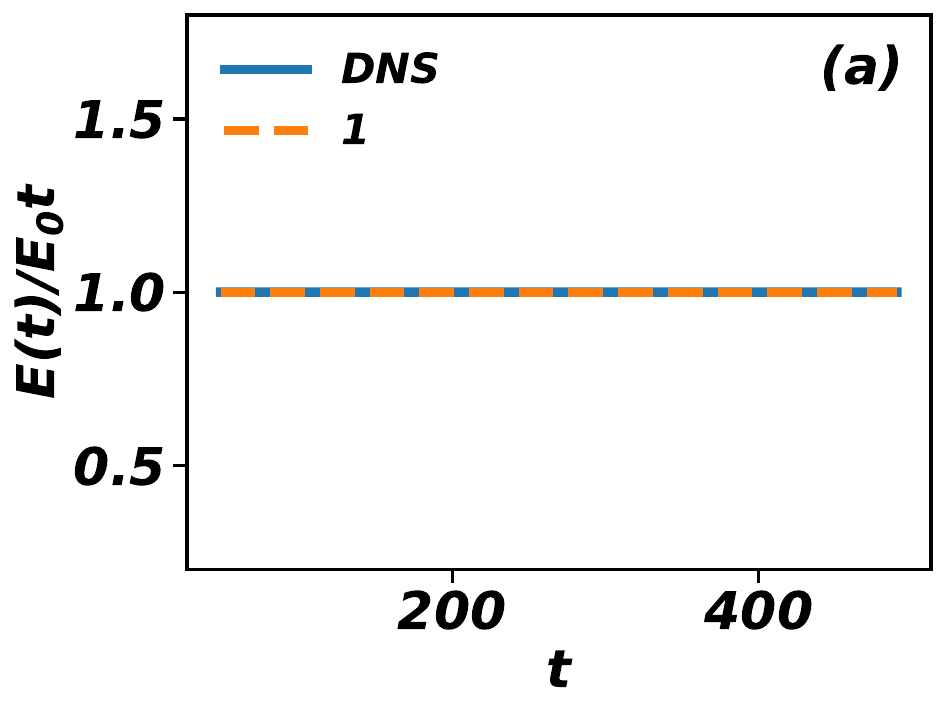}
\includegraphics[scale=0.24]{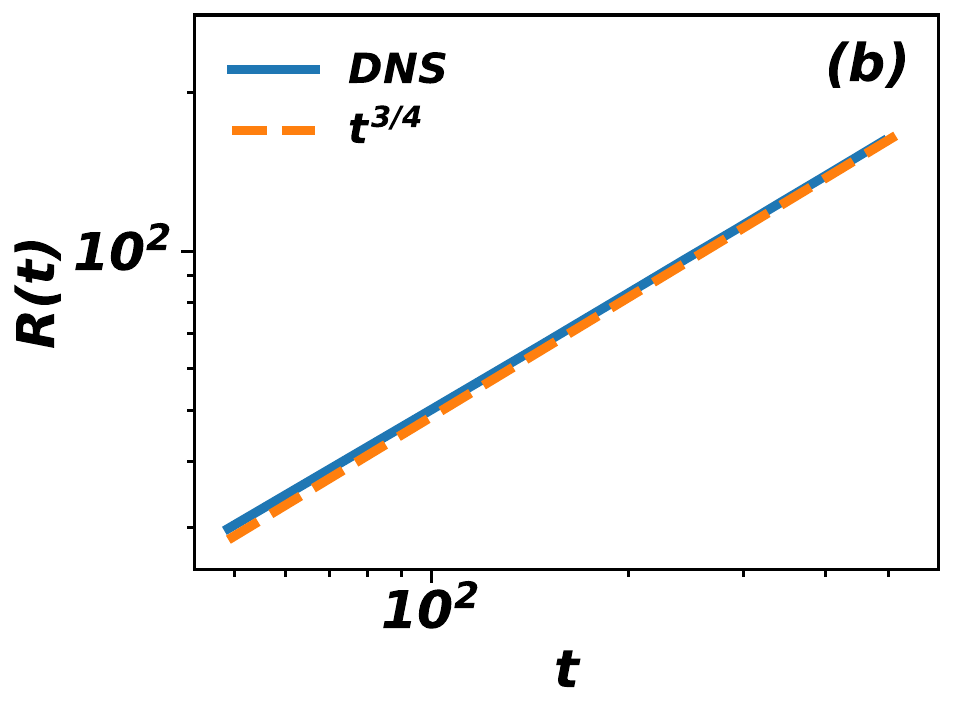}
\includegraphics[scale=0.24]{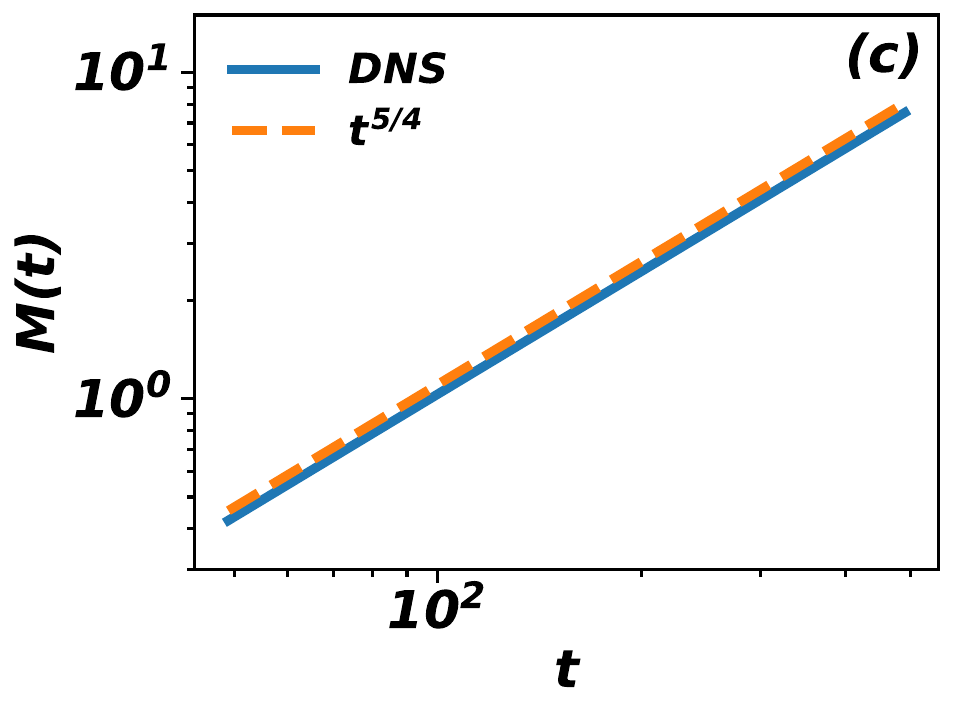}
\caption{Power law growth of (a) total energy $E(t)=E_0 t$, (b) radius of the shock front $R(t)\sim t^{3/4}$, and (c) total radial momentum $M(t)\sim t^{5/4}$ of the system. Solid lines represent the results from DNS, and dashed lines represent the respective power laws. The DNS data shown here are for ambient gas density $\rho_0=0.382$, $A_0=10^{-4}$, $C_1=C^*_1$, $C_2=C^*_2$, and $\zeta=0$.  The results are for the central driving in two dimensions with $\delta=1$.}
\label{dnspowerscaling}
\end{figure}

We now present the parametric study of the DNS of NSE. We first benchmark the DNS using the same criteria that we used for EDMD, i.e. by validating the growth of total energy, radial momentum, and radius of shock front:   $E(t)\sim t$, $M(t)\sim t^{5/4}$ [Eq.~(\ref{eq:totalmomentum})], and $R(t)\sim t^{3/4}$ [Eq.~(\ref{eq:totalradius})], in two dimensions for $\delta=1$.  We first confirm that in the DNS, the total energy increases as $E_0 t$ as can be seen from \fref{dnspowerscaling}(a).  The DNS reproduce the power law growth for both shock radius as well as radial momentum for large times, as can be seen in \fref{dnspowerscaling}(b) and (c). 

Before comparing the DNS results with results from EDMD simulations, we first examine the role of the various parameters like EOS, dissipation coefficients on the data. We point out that we obtain data collapse of the data for different times  when appropriately scaled [see Sec.~\ref{comparisonEDMDhydro}]. For the dependence on parameters, we examine the data for one time.

\begin{figure}
\includegraphics[scale=0.4]{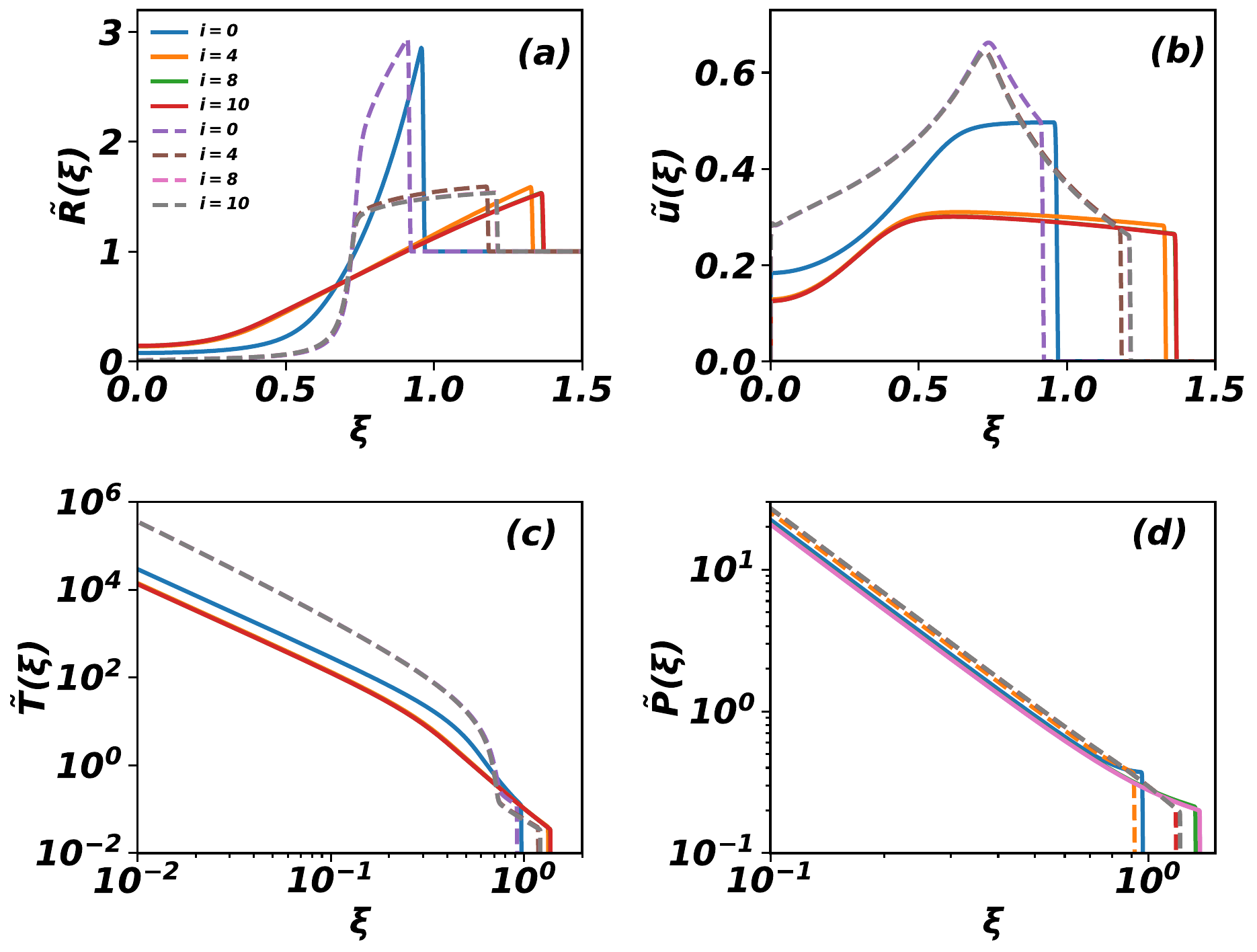}
\caption{The role of the EOS on the DNS data for (a) density $\widetilde{R}(\xi)$, (b) velocity $\widetilde{u}(\xi)$, (c) temperature $\widetilde{T}(\xi)$, and (d) pressure $\widetilde{P}(\xi)$. The virial EOS [see \eref{virialEOS}] is  truncated at $i=0$, $4$, $8$, $10$. The DNS data shown here are for initial density $\rho_0=0.382$, $A_0=10^{-4}$, $C_1=C^*_1$, $C_2=C^*_2$, $\zeta=0$, and time $t=2t'_0$, where $t'_0=489.1$. The dashed lines are for central driving and solid lines are for uniform driving in two dimensions.}
\label{dnsvirial}
\end{figure}

In \fref{dnsvirial}, we show the variation of non-dimensionalised thermodynamic functions, obtained from the DNS of NSE [Eqs.~(\ref{nse_mass})--(\ref{nse_energy})], with $\xi$ for the virial EOS with the series truncated  at the $i=0$, $4$, $8$, $10$ term. The first thing that we notice is that, for the central driving, the scaled solution of the Navier-Stokes equation covers the entire region of disturbance unlike the Euler equation. The results corresponding to $i=0$ represent the DNS for ideal EOS. The data corresponding to $i=8, 10$, lie on top of each other, thus showing negligible truncation error beyond the $10$-th term. We will therefore work with virial EOS of $10$ terms.

\begin{figure}
\includegraphics[scale=0.4]{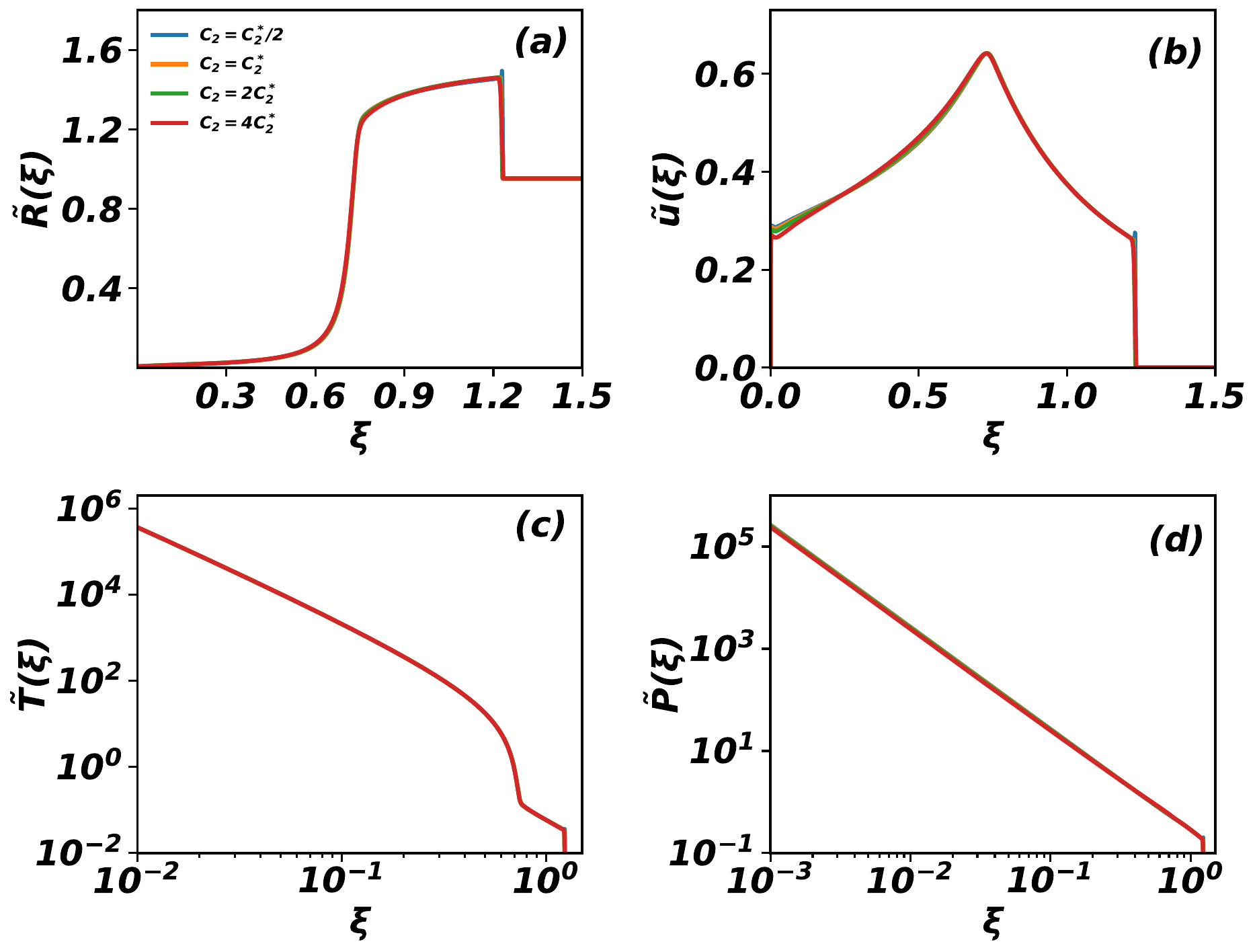}
\caption{The non-dimensionalised (a) density $\widetilde{R}(\xi)$, (b) velocity $\widetilde{u}(\xi)$, (c) temperature $\widetilde{T}(\xi)$, and (d) pressure $\widetilde{P}(\xi)$ obtained from the DNS of Eqs.~(\ref{nse_mass})--(\ref{nse_energy}) for  four different values of coefficient of viscosity $C^*_1/2$, $C^*_1$, $2C^*_1$, $4C^*_1$, keeping coefficient of heat conduction fixed at $C_2=C^*_2$. The DNS data shown here are for initial density $\rho_0=0.382$, $A_0=10^{-4}$, virial EOS up to $10^{th}$ terms, $\zeta=0$, and time $t=2t'_0$, where $t'_0=489.1$.  The results are for the central driving in two dimensions.}
\label{dnsmu}
\end{figure}

To study the role of viscosity, we study the DNS with four different values of coefficient of viscosity $C^*_1/2$, $C^*_1$, $2C^*_1$, $4C^*_1$ keeping the heat conduction fixed at $C_2=C^*_2$. We find that the value of $C_1$ does not affect the results much as can be seen from \fref{dnsmu}, where the different thermodynamics quantities are shown. We conclude that the DNS data are not sensitive to the value of the viscosity of the gas. 

\begin{figure}
\includegraphics[scale=0.4]{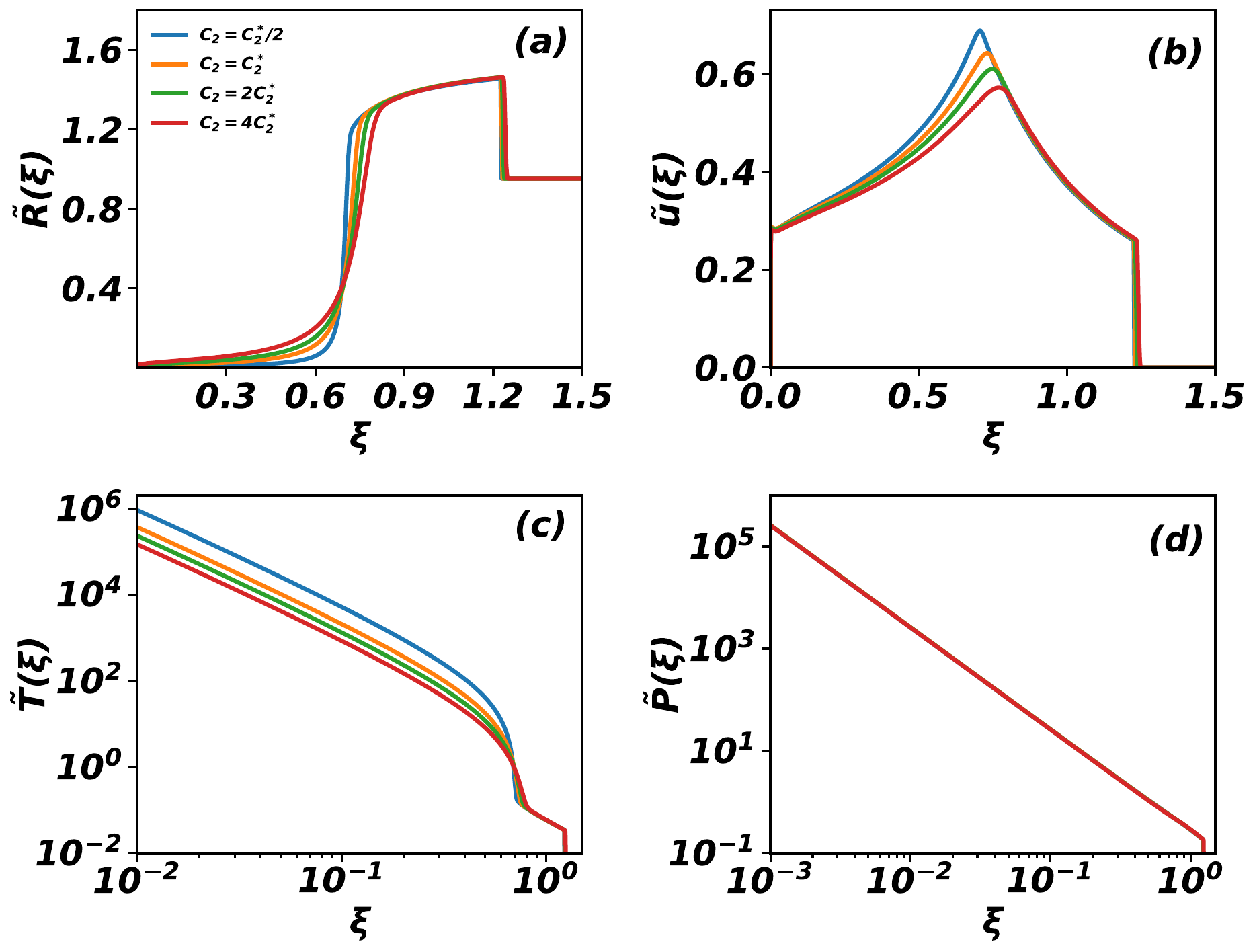}
\caption{The non-dimensionalised  (a) density $\widetilde{R}(\xi)$, (b) velocity $\widetilde{u}(\xi)$, (c) temperature $\widetilde{T}(\xi)$, and (d) pressure $\widetilde{P}(\xi)$ obtained from the DNS of Eqs.~(\ref{nse_mass})--(\ref{nse_energy}) for  four different values of coefficient of heat conduction $C^*_2/2$, $C^*_2$, $2C^*_2$, $4C^*_2$, keeping coefficient of viscosity fixed at $C_1=C^*_1$. The DNS data shown here are for initial density $\rho_0=0.382$, $A_0=10^{-4}$, virial EOS up to $10^{th}$ terms, $\zeta=0$, and time $t=2t'_0$, where $t'_0=489.1$.   The results are for the central driving in two dimensions.}
\label{dnseta}
\end{figure}

To study the role of heat dissipation, we study the DNS with  four different values of coefficient of heat conductivity $C^*_2/2$, $C^*_2$, $2C^*_2$, $4C^*_2$ keeping viscosity $C_1=C^*_1$ fixed. Unlike the case of viscosity, we find that the different thermodynamic quantities, except pressure, depend on the value of $C_2$, as can be seen from \fref{dnseta}.  

\section{Results : comparison between the Euler equation, EDMD, and DNS} \label{comparisonEDMDhydro}

We now compare the results from the different schemes that we have used to study continuous shock: simulations of discrete hard spheres using EDMD, solution of Euler equation, and DNS of the Navier-Stokes equation  for central as well as uniform driving in two dimensions, and for central driving in three dimensions. 

\begin{figure}
\includegraphics[scale=0.4]{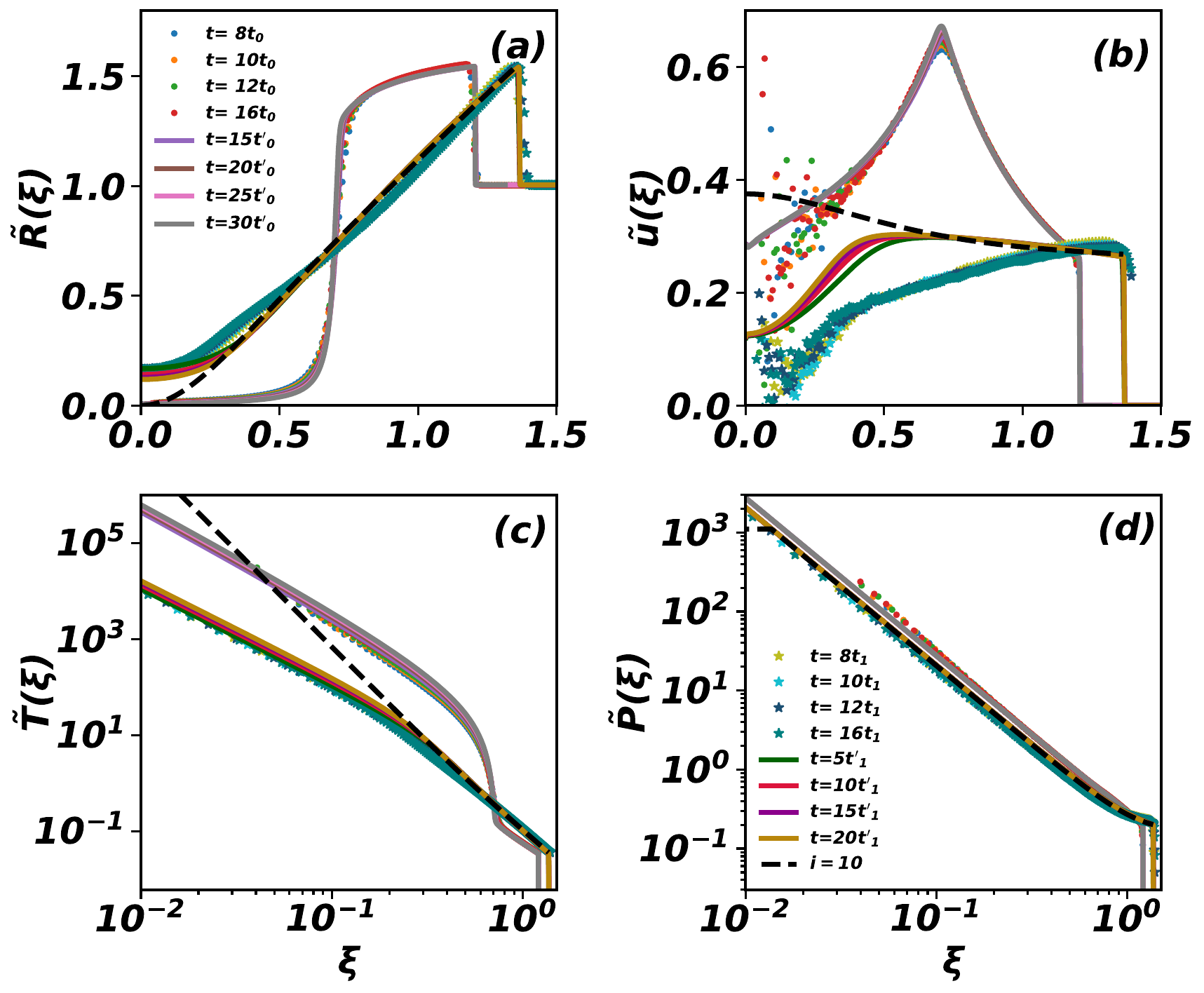}
\caption{The comparison between the profiles of non-dimensionalised  (a) density $\widetilde{R}(\xi)$, (b) velocity $\widetilde{u}(\xi)$, (c) temperature $\widetilde{T}(\xi)$, and (d) pressure $\widetilde{P}(\xi)$ obtained from Euler Eqs.~(\ref{virialmass})--(\ref{virialenergy}), EDMD, and the DNS of Navier-Stokes Eqs.~(\ref{nse_mass})--(\ref{nse_energy}) in two dimensions. The EDMD data at for different times $t=8t_0$, $10t_0$, $12t_0$, $16t_0$ for central driving (Dots), and at $t=8t_1$, $10t_1$, $12t_1$, $16t_1$ for uniform driving (Stars), where $t_0,t_1=1000$. The solid lines represent the DNS of Navier-Stokes equation at four different times $t=15t'_0$, $20t'_0$, $25t'_0$, $30t'_0$ for central driving, and at $t=5t'_1$, $10t'_1$, $15t'_1$, $20t'_1$ for uniform driving, where $t'_0,t'_1=489.1$. The dashed lines represent the results of Euler equation for uniform driving. The data shown here for DNS are for initial density $\rho_0=0.382$, $A_0=10^{-4}$, virial EOS up to $10^{th}$ terms, $\zeta=0$, $C_1=C^*_1$, $C_2=C^*_2$, and the data for EDMD are for ambient gas density $\rho_0=0.382$, $E_0=3.3\times 10^{-4}$, $R_0=30.0$, $\Delta t=1.1$, and $8\times 10^6$ number of hard sphere particles.} 
\label{tvnsedmddns}
\end{figure}

We first show the results in two dimensions. The different  non-dimensionalised quantities for four different times are shown in \fref{tvnsedmddns} for both uniform and central driving. We first note the results of DNS for different times collapse onto a single curve verifying the scaling Eqs.~(\ref{rescadist})--(\ref{rescapres}). The DNS data are able to capture the EDMD data for $\widetilde{R}(\xi)$,  $\widetilde{T}(\xi)$, and $\widetilde{P}(\xi)$. Interestingly, we find that the power law behavior of thermodynamic quantities from the DNS as well as the EDMD in both the drivings are the same. For the velocity field $\widetilde{u}(\xi)$ for the uniform driving, the results from DNS, while matching with the EDMD results near the shock front, has a quantitative mismatch away from the shock center, as can be seen from \fref{tvnsedmddns}(b). A possible reason for this mismatch could be that, in the EDMD simulations with uniform driving, there is a non-negligible non-radial velocity at all distances. 

For uniform driving, the qualitative behavior of scaling functions for hard sphere gas, obtained from the numerical solution of Euler equation, are exactly same as the exact solution for ideal gas.  In fact, the variation $\widetilde{T}\to \xi^{-2}$, in EDMD, indicates that $T(r,t)\to r^0 t^{-1/2}$ close to the shock center, which means that the temperature $T(r,t)$ decreases in time and the slope with respect radial distance $r$ is zero, while the behavior $\widetilde{T}\to \xi^{-4}$ in numerical solution shows that the temperature $T(r,t)$ varies as $r^{-2} t^{1}$ which gives divergent spatial slope of temperature at the shock center. The divergent temperature leads to infinite energy in the system, which is unphysical for a finitely driven system. Since the heat conduction term put a boundary condition: $\vec{\nabla}T=0$,  the solution of full Navier-Stokes equation  resolve the discrepancy between EDMD and hydrodynamics.

\begin{figure}
\includegraphics[scale=0.4]{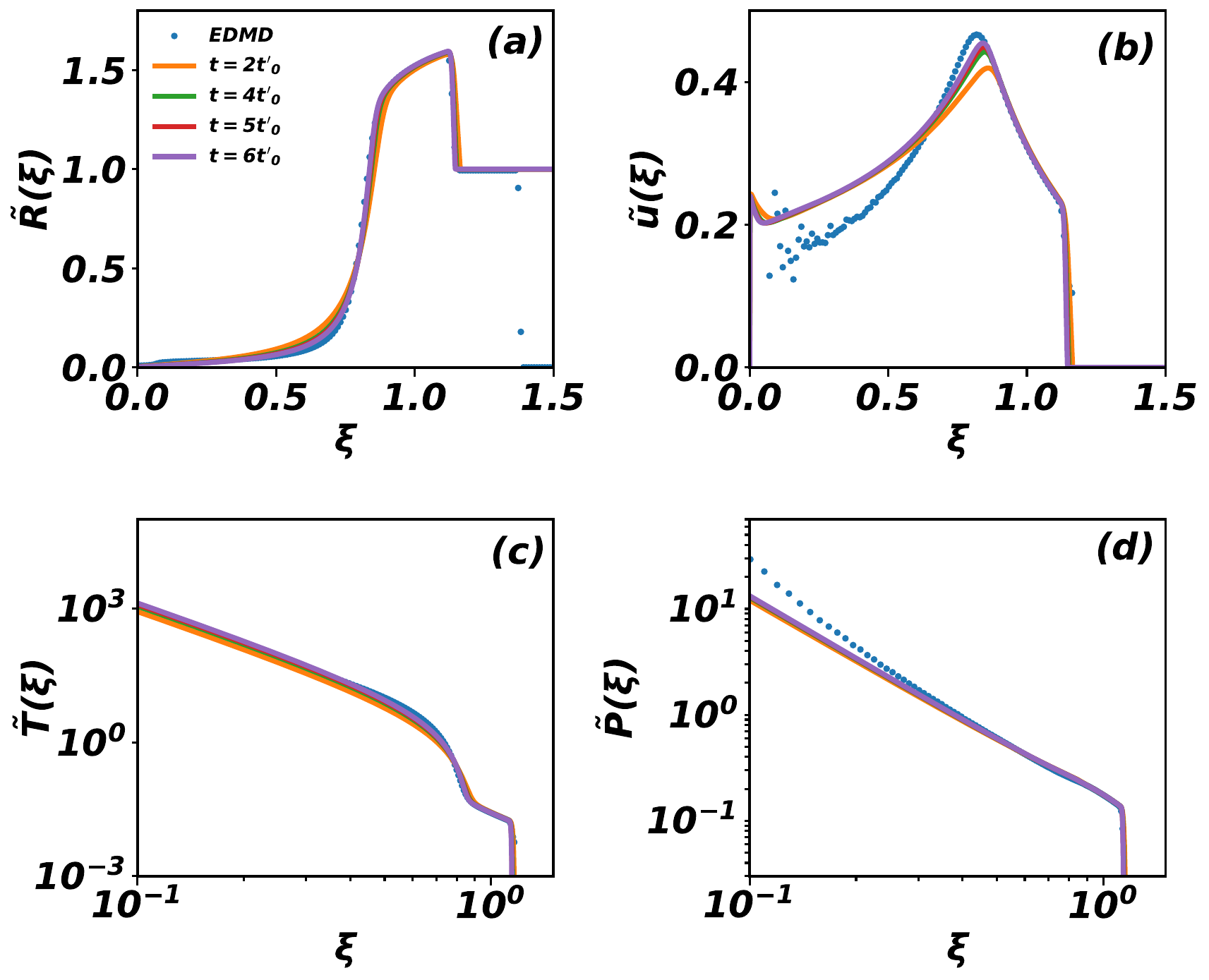}
\caption{The comparison between the profiles of non-dimensionalised  (a) density $\widetilde{R}(\xi)$, (b) velocity $\widetilde{u}(\xi)$, (c) temperature $\widetilde{T}(\xi)$, and (d) pressure $\widetilde{P}(\xi)$ obtained from EDMD, and the DNS of Navier-Stokes Eqs.~(\ref{nse_mass})--(\ref{nse_energy}) in three dimensions. The symbols represent single time EDMD data at time $t= 3t_0$, where $t_0=1357.2$, and the solid lines denote the results of DNS at four different times $t=2t'_0$, $4t'_0$, $5t'_0$, $6t'_0$ where $t'_0=69.2$. The data shown here for DNS are for initial density $\rho_0=0.4013$, $A_0=10^{-4}$, virial EOS up to $10^{th}$ terms, $\zeta=0$, $C_1=10C^*_1$, $C_2=8.35C^*_2$, and the data for EDMD are for ambient gas density $\rho_0=0.4013$, $E_0=2.5\times 10^{-6}$, $R_0=15.0$, $\Delta t=0.15$, and $4\times 10^7$ number of hard sphere particles.} 
\label{tvnsedmddns3d}
\end{figure}

We now discuss the results in three dimensions for central driving. The different non-dimensionalised quantities for four different times are shown in \fref{tvnsedmddns3d}. To match with the EDMD data, we have run the DNS for different values of $C_1$ and $C_2$, and chosen values with the best match. To do so, we started with the initial values of $C_1$, and $C_2$ as $C^*_1$, and $C^*_2$ respectively and then we increased these values systematically till we obtained a best fit (visually) to the data for all the scaling. These results of central driving in three dimensions are very similar to the results in two dimensions, and we conclude that the NSE are able to provide the correct hydrodynamics for the driven shock problem. 

\section{Summary and discussion} \label{summary}

In summary, we studied the hydrodynamics  of shocks in a gas in which energy is continuously input ($E(t) \sim t^\delta$) either at one localized region in space (central driving) or throughout the growing affected region in time (uniform driving). Different schemes were used to study this problem:  EDMD simulations, DNS of Navier-Stokes equation, and numerical solution of the Euler equation.  

For uniform driving, we showed that the power law exponents of thermodynamic quantities, obtained from the solution of Euler equation, are independent of $\delta$. We showed that, for uniform driving, the solution of Euler equation does not match with the EDMD data close to the shock center in terms of different power law exponents, while it matched near the shock front. Inclusion of dissipation terms in terms of the Navier-Stokes equation is able to describe the simulation results near the shock center also.

For central driving, we showed that the self-similar solution of the Euler equation is singular and does not extend over the full spatial region of the shock.  However, the numerical solutions of Navier-Stokes equation are able to produce self-similar solution that extend all the way to the shock center, thus  showing the necessity of dissipation terms to even have a sensible solution. The Navier-Stokes equation is also able to describe the simulation results for the different thermodynamic quantities, provided the heat conduction and viscosity are chosen parametrically. 

We conclude that even though the continuous drive takes the system far from equilibrium, Navier-Stokes equation continues to give a good description of the system. For uniform driving, the reason why Euler equation does not provide a good hydrodynamic description remains the same as that for the single impact. Temperature diverges at the shock center, within the Euler equation. Adding a heat conduction term regularizes this behavior with the radial derivative going to zero. This is what is observed in EDMD simulations also. Thus, the solution of the Euler equation does not respect the boundary conditions seen in simulations, leading to an incorrect description. For central driving, the Euler equation fails poorly at describing the system.

Incorporating heat conduction in the continuity equations altered the scaling near the shock center for the case of single impact. This crossover has been quantified in earlier work~\cite{ganapa2021blast,chakraborti2021blast,singh2023blast}. Generalizing these results to the case of driven shock is an interesting problem for future research. However, obtaining clean data near the shock center is a more challenging problem for driven shocks. Central driving introduces a new length scale, defined by the region of driving, and hence taking the $r\to 0$ limit requires simulations of much larger systems. However, it will be easier to study the crossover in the case of uniform driving.

Shock propagation has also been studied in granular systems where collisions between particles are inelastic. The creation of a shock due to a single impact is relevant for the study of crater formation due to the  impact of particles having a high initial energy~\cite{grasselli2001crater}, dropping a steel ball vertically into a container of small glass beads~\cite{walsh2003morphology}, or due to the single impact of steel ball on flowing glass beads~\cite{boudet2009blast}, and has been studied using scaling and simulations particle based models~\cite{jabeen2010universal,pathak2012shock}. For studying such shocks, the the TvNS theory has been  modified for dissipative systems~\cite{barbier2016microscopic,barbier2015blast}. Shocks due to continuous driving are also relevant for granular systems. For example,  granular fingering or pattern formation due to continuous injection of a viscous liquid in dry dense granular medium~\cite{cheng2008towards,sandnes2007labyrinth,pinto2007granular,johnsen2006pattern,huang2012granular}, impinging of gas jets vertically on a granular bed~\cite{metzger2009craters} create outwardly moving disturbances and have been studied using scaling and particle based models~\cite{joy2017shock}. Generalizing the theory~\cite{barbier2016microscopic,barbier2015blast}  to continuous driving, and checking its validity with experiments and simulation is a promising area of future research.

\begin{acknowledgements}
The simulations were carried out on the supercomputer Nandadevi at The Institute of Mathematical Sciences.
\end{acknowledgements}

\section*{Data availability} The datasets generated during the current study are available from the corresponding author on reasonable request.

\appendix

\section{\label{appendixA} Review of exact solution of Euler equation}

In this Appendix, we provide the analytical solution of ordinary differential Eqs.~(\ref{idealmass})--(\ref{idealenergy}) satisfying the  Rankine-Hugoniot boundary conditions (see Eqs.~(\ref{idealRHmass})--(\ref{idealRHenergy})) for non-dimensionalised scaling functions $\widetilde{R}$, $\widetilde{u}$, $\widetilde{T}$, and $\widetilde{P}$ as defined in Eqs.~(\ref{rescadist})--(\ref{rescapres}). On further simplifying the Eqs.~(\ref{idealmass})--(\ref{idealenergy}) for $d\log \xi /d\widetilde{u}$ and $d\log \widetilde{R}/d\log \xi$, we obtain
\begin{align}
&\frac{d\widetilde{u}}{d\log \xi} =\frac{\widetilde{u}(\gamma \widetilde{u}-\alpha)[10\alpha^2+2\widetilde{u}[(2+\delta)( \widetilde{u} - 2\alpha)   +\gamma\alpha(\delta-3)]+3\widetilde{u}^2(2+\delta)(\gamma-1)]}{ (2+\delta)[2\alpha^3 - 2\alpha\widetilde{u}(\gamma+2)+\alpha\widetilde{u}^2(3+2\gamma+\gamma^2)  - 2\gamma \widetilde{u}^3- \widetilde{u}^2(\gamma-1)(\gamma \widetilde{u}-\alpha)]}, \label{appenidealmasssimplified}\\
&\frac{d \log \widetilde{R}}{d \log \xi}= \nonumber \\
&\frac{3\widetilde{u}(2+\delta)[2\alpha^3 - 2\alpha^2\widetilde{u}(\gamma+2)+\alpha\widetilde{u}^2(3+2\gamma+\gamma^2)  - 2\gamma \widetilde{u}^3]  + 2\widetilde{u}(\gamma \widetilde{u}-\alpha)[5\alpha^2+\widetilde{u}[(2+\delta)( \widetilde{u} - 2\alpha)   +\alpha\gamma(\delta-3)]]}{(2+\delta)(\alpha-\widetilde{u}) [2\alpha^3 - 2\alpha^2\widetilde{u}(\gamma+2)+\alpha\widetilde{u}^2(3+2\gamma+\gamma^2)  - 2\gamma \widetilde{u}^3 - \widetilde{u}^2(\gamma-1)(\gamma \widetilde{u}-\alpha)]},\label{appenidealmomentumsimplified}\\
&\widetilde{T} = \frac{\widetilde{u}^2(\alpha-\widetilde{u})(\gamma-1)}{2(\gamma \widetilde{u}-\alpha))}.\label{appenidealenergysimplified}
\end{align}

The above ordinary differential equations  can be solved analytically. The solution depends on the sign of the parameter $a_1=(\gamma -2)^2 \left(\delta^2+9\right)-\left(6 \gamma ^2+26 \gamma -26\right) \delta$. When $a_1 \geq 0$, we find 
\begin{align}
&\widetilde{R}(\widetilde{u})=\frac{\alpha b_1}{\alpha-\widetilde{u}} \left(f_2(\widetilde{u})\right)^{a_8} \left(\frac{\gamma  \widetilde{u}-\alpha}{\alpha(\gamma-1)}\right)^{a_9} \exp \left(\frac{a_7}{\sqrt{a_1}} \left[\tanh ^{-1}\left(\frac{f_1(\widetilde{u})}{\sqrt{a_1}}\right)-\tanh ^{-1}\left(\frac{a_5}{\sqrt{a_1}}\right)\right]\right), \label{appenpos_g}\\
&\left(\frac{10}{2+\delta}\right)\log{\left(\frac{\xi (\widetilde{u})}{\xi_f}\right)}= \frac{a_2}{\sqrt{a_1}} \left[\tanh^{-1}\left(\frac{a_5}{\sqrt{a_1}}\right)-\tanh ^{-1}\left(\frac{f_1(\widetilde{u})}{\sqrt{a_1}}\right)\right] +a_3 \log \left(\frac{\gamma  \widetilde{u}-\alpha}{\alpha(\gamma-1)}\right) + \log \left(\frac{(\alpha^2f_2(\widetilde{u}))^{a_4}}{\widetilde{u}^2}\right)+b_2.\label{appenpos_xi}
\end{align}

On the other hand, when $a_1 <0$, we obtain
\begin{align}
&\widetilde{R}(\widetilde{u})=\frac{\alpha b_1}{\alpha-\widetilde{u}} \left(f_2(\widetilde{u})\right)^{a_8} \left(\frac{\gamma  \widetilde{u}-\alpha}{\alpha(\gamma-1)}\right)^{a_9} \exp \left(\frac{a_7}{\sqrt{-a_1}} \left[\tan ^{-1}\left(\frac{a_5}{\sqrt{-a_1}}\right)-\tan ^{-1}\left(\frac{f_1(\widetilde{u})}{\sqrt{-a_1}}\right)\right]\right), \label{appenneg_g}\\
&\left(\frac{10}{2+\delta}\right)\log{\left(\frac{\xi (\widetilde{u})}{\xi_f}\right)}= \frac{a_2}{\sqrt{-a_1}} \left[\tan ^{-1}\left(\frac{f_1(\widetilde{u})}{\sqrt{-a_1}}\right)-\tan^{-1}\left(\frac{a_5}{\sqrt{-a_1}}\right)\right] +a_3 \log \left(\frac{\gamma  \widetilde{u}-\alpha}{\alpha(\gamma-1)}\right)+ \log \left(\frac{(\alpha^2f_2(\widetilde{u}))^{a_4}}{\widetilde{u}^2}\right)+b_2, \label{appenneg_xi}
\end{align}
where,
\begin{align}
&a_2=\frac{2 \left( \left(6 \gamma ^3-11 \gamma ^2-3 \gamma +2\right) \delta^2+\left(19 \gamma ^3+16 \gamma ^2+33 \gamma -32\right) \delta + 39 \gamma ^3-99 \gamma ^2+78 \gamma-72\right)}{ (2 \gamma +1) (3 \gamma -1) (\delta+2)}, \\
&a_3=\frac{10 (\gamma -1)}{(2 \gamma +1) (\delta+2)}, \\
&a_4=\frac{ \left(6 \gamma ^2+\gamma -1\right) \delta -13 \gamma ^2 +7 \gamma-12}{(2 \gamma +1) (3 \gamma -1) (\delta+2)}, \\
&a_5=\frac{ \left(\gamma ^2+5 \gamma -4\right) \delta-3 \gamma ^2+5 \gamma-8}{\gamma +1}, \\
&a_6=\frac{2 \left( (2 \gamma ^2 +4 \gamma  -6) \delta-\gamma ^2+8 \gamma-7\right)}{(\gamma +1)^2},\\
&a_{7}=\frac{6 (\gamma +3) \left(\left(\gamma ^2+\gamma -1\right) \delta-3 \gamma ^2+2 \gamma -2\right)}{ \left(6 \gamma ^2+\gamma -1\right)}, \\
&a_{8}=\frac{3 \left(\gamma ^2+1\right)}{6 \gamma ^2+\gamma -1},\\
&a_{9}=\frac{3}{2 \gamma +1}, \\
&b_1=\left(\gamma +1\right)^{a_9} a_6^{-a_8}, \\
&b_2=a_3 \log \left({\gamma +1}\right)-a_4 \log (a_6)+2 \log \left(\frac{2}{\gamma +1}\right), \\
&f_1(\widetilde{u})=(3 \gamma -1)(\delta+2) \widetilde{u}/\alpha + (\gamma -2)\delta-3 \gamma-4, \\
&f_2(\widetilde{u})=(3 \gamma -1) (\delta+2) \widetilde{u}^2/\alpha^2+2  ((\gamma -2) \delta-3 \gamma -4)\widetilde{u}/\alpha+10.
\end{align}
We have checked for the correctness of the solution by checking that they match with the numerical solution of the differential equations.

The value of $\xi_f$ can be obtained by using the energy constraint $E(t)=E_0t^\delta$ [see \eref{idealbeta}]. The analytical solutions show the following asymptotic behavior of scaling functions, as $\xi\to 0$,
\begin{align}
\widetilde{u}-\frac{\alpha}{\gamma}&\to \xi^{\frac{2\gamma+1}{\gamma-1}} \label{appenasy_V}\\
\widetilde{R} &\to \xi^\frac{3}{\gamma-1} \label{appenasy_G}\\
\widetilde{T} &\to \xi^{-\frac{2\gamma+1}{\gamma-1}} \label{appenasy_Z}. 
\end{align}


\end{document}